\documentclass{aastex}
\usepackage{pslatex}
\usepackage{emulateapj5}
\usepackage{epsf}

\slugcomment{{\it The Astrophysical Journal Supplement, 2003, 145} }
\newcommand{\uvolt}{\mbox{\ensuremath{\mu\textrm{V}}}}
\newcommand{\mvolt}{\mbox{\ensuremath{\textrm{mV}}}}
\newcommand{\nvolt}{\mbox{\ensuremath{\textrm{nV}}}}
\newcommand{\nA}{\mbox{\ensuremath{\textrm{nA}}}}
\newcommand{\mA}{\mbox{\ensuremath{\textrm{mA}}}}
\newcommand{\ms}{\mbox{\ensuremath{\textrm{ms}}}}

\newcommand{\Hz}{\mbox{\ensuremath{\textrm{Hz}}}}
\newcommand{\kHz}{\mbox{\ensuremath{\textrm{kHz}}}}

\newcommand{\GHz}{\mbox{\ensuremath{\textrm{GHz}}}}
\newcommand{\rHz}{\mbox{\ensuremath{\textrm{Hz}^{-1/2}}}}
\newcommand{\cm}{\mbox{\ensuremath{\textrm{cm}}}}
\newcommand{\rms}{\mbox{\ensuremath{\textrm{RMS}}}}
\newcommand{\ukelvin}{\mbox{\ensuremath{\mu\textrm{K}}}}
\newcommand{\ddeg}{\mbox{\ensuremath{{\rlap.}^\circ}}}
\shorttitle{MAP Radiometers}
\shortauthors{Jarosik et al.}
\begin{document}

\title{Design, Implementation and Testing of the {\sl MAP} Radiometers}
\author{N. Jarosik \altaffilmark{1}, 
C. L. Bennett \altaffilmark{2},
M. Halpern \altaffilmark{3},
G. Hinshaw \altaffilmark{2}, \\
A. Kogut \altaffilmark{2}, 
M. Limon \altaffilmark{2}, 
S. S. Meyer \altaffilmark{4},
L. Page \altaffilmark{1},
M. Pospieszalski \altaffilmark{9},\\
D. N. Spergel \altaffilmark{5},
G. S. Tucker \altaffilmark{6},
D. T. Wilkinson \altaffilmark{1},
E. Wollack \altaffilmark{2},\\
E. L. Wright \altaffilmark{7},
Z. Zhang \altaffilmark{8}}

\altaffiltext{1}{Dept. of Physics, Jadwin Hall, Princeton, NJ 08544}
\altaffiltext{2}{Code 685, Goddard Space Flight Center, 
                 Greenbelt, MD 20771}
\altaffiltext{3}{Dept. of Physics, Univ. Brit. Col., Vancouver, B.C., Canada V6T 1Z4}
\altaffiltext{4}{Astronomy and Physics, University of Chicago, 5640 South Ellis Street, 
                 LASP 209, Chicago, IL 60637}
\altaffiltext{5}{Dept of Astrophysical Sciences, Princeton University,
                 Princeton, NJ 08544}
\altaffiltext{6}{Dept. of Physics, Brown University, Providence, RI 02912}
\altaffiltext{7}{Astronomy Dept., UCLA, Los Angeles, CA 90095}
\altaffiltext{8}{Code 555, Goddard Space Flight Center, 
                 Greenbelt, MD 20771}
\altaffiltext{9}{National Radio Astronomy Observatory, Charlottesville VA, 
                 22903-2475}
\email{jarosik@pupgg.princeton.edu}

\begin{abstract}
The Microwave Anisotropy Probe ({\sl MAP}) satellite,  launched June 30, 2001, 
will produce full sky maps of the cosmic microwave background radiation in 5
frequency bands spanning
20 - 106~\GHz. {\sl MAP} contains 20 differential radiometers built with
High Electron Mobility Transistor (HEMT) amplifiers
with passively cooled input stages. 
The design and test techniques used to evaluate and minimize systematic errors
and the pre-launch performance of the radiometers for all five bands 
are presented.
\end{abstract}


\section{Introduction}
This paper describes the design, implementation and testing of the 
microwave radiometers and support electronics incorporated in the 
Microwave Anisotropy Probe ({\sl MAP})
satellite.
The {\sl MAP} mission is the second  
in NASA's Medium-class Explorers (MIDEX) program \citep{missionpaper}.
Its purpose is to produce multi-frequency full sky maps of 
the microwave sky, from which maps of the cosmic microwave
background radiation (CMB) can be extracted. 
The design of {\sl MAP} is the result of a detailed study
which involved balancing the mission's three major 
interdependent performance requirements, angular
resolution, sensitivity and systematic error suppression, 
while remaining consistent with   the resources provided by MIDEX program.
Some of the major design parameters which resulted are:
angular resolution of   $\approx 0 \ddeg 25$;
minimally correlated pixel noise;
full sky coverage; frequency coverage of $20-106\;\GHz$  \ in 5 bands;
polarization sensitivity; calibration accuracy  better than $0.5\;\%$; 
sensitivity of $\approx 30\;\ukelvin$ per  $3.2 \times 10^{-5}$ sr/pixel.

The {\sl MAP} instrument is composed of 20 differential HEMT based radiometers coupled to
the sky by a set of back-to-back off axis Gregorian optics. The optics and input stages
of the radiometers are passively cooled by radiation to space. The angle subtended between
the two input beams of each radiometer is $\approx 140^\circ$. {\sl MAP} observes from the
second Earth-Sun Lagrange point using a compound spin motion producing highly interconnected
scan patterns on the sky. These patterns greatly facilitate reconstruction of the temperature
maps from the measured differential signals while suppressing systematic errors.
Details of the entire mission, including diagrams of spacecraft and instrument layout and
estimated uncertainties in the measured angular power spectra
are presented in \citet{missionpaper}. 

Control of possible sources of systematic errors (features in the maps, other than
 random noise, which are not present on the sky)
  was a high priority and greatly influenced many aspects of {\sl MAP}'s design.
Systematic errors
 can arise from  several sources, including sidelobe pickup from foreground objects (the Sun, Moon, Earth and Galaxy),
 instrument instabilities (amplifier gain fluctuations, 
thermal drifts, etc.) or the data processing used to 
convert the time ordered data into calibrated sky maps.
Careful optical
design \citep{opticspaper, feedspaper} orbit selection, and
 a scan strategy  that keeps most contaminating sources far from the
lines of sight of the instrument were used to minimize sidelobe pickup of foreground objects . 
The data processing procedure  was checked
 for possible systematic errors by processing
simulated data sets through the entire data processing pipeline and comparing the 
outputs to the simulation inputs. 

Instrumental effects include instabilities
inherent in the radiometers and those caused by environmental changes to the
instrument. 
Intrinsic radiometer instabilities were minimized by the use of a differential
radiometer design, the details of which are  presented later in this paper. 
The entire observatory, comprising the instrument and spacecraft structure, was
carefully designed to maintain as stable an environment as possible for the instrument.
Specific examples are the scan strategy, which maintains a constant angle of insolation
on the observatory, thermal isolation of the instrument from
 the spacecraft bus, and the nearly constant spacecraft power dissipation, all of which
minimize thermal disturbances which could influence instrument performance.

A systematic error budget was established before the instrument build and estimates of
 contributions from all known potential
signal sources were tracked throughout the program. Signal sources
 were classified as either spin synchronous or random. Random
signals are those which are uncorrelated with orientation of the observatory.
 Random signals tend to average to zero, only
the residual terms contribute to errors in the final maps.
Spin synchronous
signals, such as Galactic pickup from side-lobes or radiometer drifts due to
 spin synchronous thermal fluctuations do not average to zero,
and therefore must be limited to much smaller values in the time ordered data. 
{\sl MAP} is designed to limit the magnitude of systematic errors 
signals to less
 than $4.5$~\ukelvin\ \rms\ in the uncorrected sky maps. Simulations
indicate that spin synchronous artifacts in
 the time ordered data are suppressed by a factors of
2 or more by the map making algorithm given our scan pattern, so the corresponding allowed
 spin synchronous systematic error level in the time stream
is $9$~\ukelvin. Of this a $2.6$~\ukelvin\ contribution is budgeted for emission
 from cryogenic components in the feed horns and optical system,
 $2.8$~\ukelvin\ from
 the radiometers, data collection system, and supporting electronics, with 
the remainder allocated to emission sources external to the
satellite picked up by the side-lobe response of the optical system.

The {\sl MAP} radiometers were designed to meet the systematic error
requirement \emph{without applying any corrections} to the radiometric data.
The ultimate goal of these efforts is to produce  high-quality sky maps
with negligible striping along the scan directions and
 nearly diagonal pixel-pixel noise correlation matrices. 

\section{Radiometer Design}
\subsection{Overview} \label{sec:rad_overview}
{\sl MAP} contains 20 differential radiometers covering 5 frequency bands. They are direct conversion
 radiometers (no mixing to an intermediate frequency) and share the same
basic design. Operation of the radiometers is similar to the continuous comparison radiometer
described by \citet{predmore}. Amplification is provided by High Electron Mobility Transistor (HEMT)
amplifiers, with the input stages passively cooled to $\approx 90$~K to lower
the system noise. The total microwave gain of each radiometer was selected so that
the power input to each diode detector would be $\approx -23$~dBm during
 science observations. The resulting signal
level is low enough to keep the diode
detectors in the square law regime, but high enough so that intrinsic detector noise, and that added
 by the  post detection electronics, degrade radiometer sensitivity by $< 0.5\%$. 

A total power radiometer's sensitivity can be described by the radiometer equation \citep{Kraus, dicke} ,
\begin{equation}
dT = T_\mathrm{sys} \sqrt{\frac{1}{\Delta \nu_\mathrm{eff} \ \tau_\mathrm{int}}  + \left( \frac{\Delta G }{G}\right)^2}, \label{eqn:rad_sense}
\end{equation}
 which relates the radiometer noise, $dT$, for an integration period
  $\tau_\mathrm{int}$, to the input referenced system
 noise temperature, $T_\mathrm{sys}$, and the effective RF bandwidth
of the radiometers, $\Delta \nu_\mathrm{eff}$, given by
\begin{equation}
\Delta \nu_\mathrm{eff} = \frac{[\int G(\nu) d\nu ]^2}{\int [G(\nu)^2]d\nu } 
\end{equation}
where $G(\nu)$  is the frequency dependent \emph{power} response of the radiometer.
The second term under the radical in equation~\ref{eqn:rad_sense}
is the fractional power gain variation of the radiometer on the time scale of the integration,
and would vanish for an ideal radiometer system. 

The low noise and wide bandwidth of HEMT amplifiers make them strong candidates
for use in the measurement of continuum signals, such as the CMB radiation.
Unfortunately these amplifiers 
exhibit small fluctuations of their microwave gain with an approximate
`$1/f$' frequency spectrum \citep{1/fpaper,corrnoisepaper} which results in 
an appreciable gain variation term as described above.
 Although these fluctuations are small, when combined with the
extremely wide microwave bandwidth of the amplifiers, they limit the 
sensitivity achievable by simple total power radiometers for integration times longer
 than several milliseconds.
The frequency at which the gain fluctuations contribute
a variance to a total power radiometer's output equal to the intrinsic radiometric noise
is parameterized as the $1/f$ knee frequency, $f_\mathrm{amp}$. \label{1/fdef}  Total power
 radiometers built with the {\sl MAP}
amplifiers with similar effective RF bandwidths would have $f_\mathrm{amp}$ values  ranging 
from $\approx 20$~\Hz\ for K-band up to $\approx 1$~\kHz\ for W-band.

 It is
 well known that differential radiometers can largely circumvent this problem
 if the differential `offset' of the radiometer, $T_\mathrm{off}$, and the magnitude of signal being
measured are both small compared to the radiometer's input-referenced system noise temperature.
 Such is the case with the {\sl MAP} radiometers,
 leading to their differential design.
For a given RF bandwidth the knee frequency for a differential radiometer is related
 to the amplifier's knee frequency by
\begin{equation}
f_\mathrm{knee} = f_\mathrm{amp} (T_\mathrm{off}/T_\mathrm{sys})^{2/\alpha}
\end{equation}
for a  $1/f^\alpha$ gain fluctuation spectrum.
Measurements performed on a prototype W-band {\sl MAP} amplifier \citep{1/fpaper} in a system 
with comparable RF bandwidth give $f_\mathrm{amp} \approx 1$~\kHz\ and
$\alpha \approx 0.96$. For $T_\mathrm{off} = 0.5$~K and $T_\mathrm{sys} = 130$~K, typical
 values for {\sl MAP} W-band radiometers,  the predicted knee
 frequency of the differential radiometer is $\approx 0.01$~\Hz. By comparison {\sl MAP}'s
spin frequency is 0.0077~\Hz. 

 A summary of the top level design parameters for radiometers in all five frequency bands
is given  in Table~\ref{table:nom_bands}. Note that the noise levels shown in this table  are $\sqrt{2}$ higher
than those calculated from equation~\ref{eqn:rad_sense} owing to the differential
 design of the {\sl MAP} radiometers.

\begin{table*}[t]
\caption{\small  {\sl MAP} Radiometer Design Parameters  }
\small{
\vbox{
\tabskip 1em plus 2em minus .5em
\halign to \hsize {#\hfil &\hfil#\hfil &\hfil#\hfil &\hfil#\hfil
                  &\hfil#\hfil &\hfil#\hfil \cr
\noalign{\smallskip\hrule\smallskip\hrule\smallskip}
{\sl MAP} Band Designation & K  & Ka & Q & V & W \cr  
$EIA$~\tablenotemark{a} \ WR~\#  &  WR$-42$  & WR$-28$ 
                               & WR$-22$ &   WR$-15$ & WR$-10$ \cr 
Radiometer frequency range (\GHz)   & $20-25$   & $28-36$ 
                                 & $35-46$ & $53-69$   & $82-106$ \cr 
Radiometer wavelength range (mm) & $12.0 - 15.0$ & $8.3 - 10.7$ & 
                                 $6.5 - 8.6$ & $4.3 - 5.7$ & $2.8 - 3.7$ \cr  
$\Delta \nu_\mathrm{eff}$ (\GHz)   & 4 & 5    & 8     & 13      & 19 \cr 
$ T_\mathrm{sys}~\tablenotemark{b}\ \ (K)$       & 29      & 39    &   59  &     92  
& 145 \cr 
Sensitivity/radiometer (mK~sec$^{1/2}$)~\tablenotemark{b}  & 0.65 
& 0.78 & 0.92 & 1.13 & 1.48 \cr 
Number of radiometers & 2 & 2 & 4 & 4 & 8 \cr
Radiometer designations & K11 K12 & Ka11 Ka12 & Q11 Q12 & V11 V12 & W11 W12 \cr 
                     &         &           & Q21 Q22 & V21 V22 & W21 W22 \cr 
                     &         &           &         &         & W31 W32 \cr
                     &         &           &         &         & W41 W42 \cr 
\noalign{\smallskip\hrule}
}}
\tablenotetext{a}{Electronic Industries Association designations 
for the waveguide band used to construct the radiometers. }
\tablenotetext{b}{Sensitivity and $T_\mathrm{sys}$ values are given in 
Rayleigh-Jeans temperatures.}
Top level radiometer design parameters as presented in the technical volume
of the {\sl MAP} MIDEX proposal. Actual radiometer performance 
data obtained during the MAP integration and test program are presented later.
} 
\label{table:nom_bands}
\end{table*}

\begin{figure*}
\epsscale{1.5}
\plotone{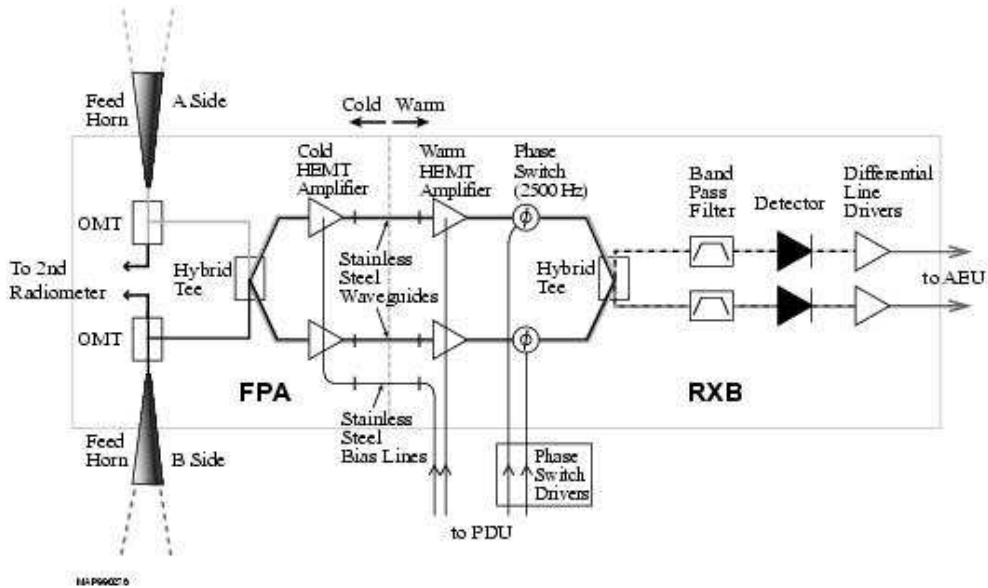}
\caption{Layout of an individual  {\sl MAP} radiometer. Components on the cold (left) side  of the
stainless steel waveguides are located in the FPA, and are passively cooled to 90~K in flight. }
\label{fig:full_rad_dia}
\end{figure*}

\begin{figure*}
\epsscale{2.0}
\plotone{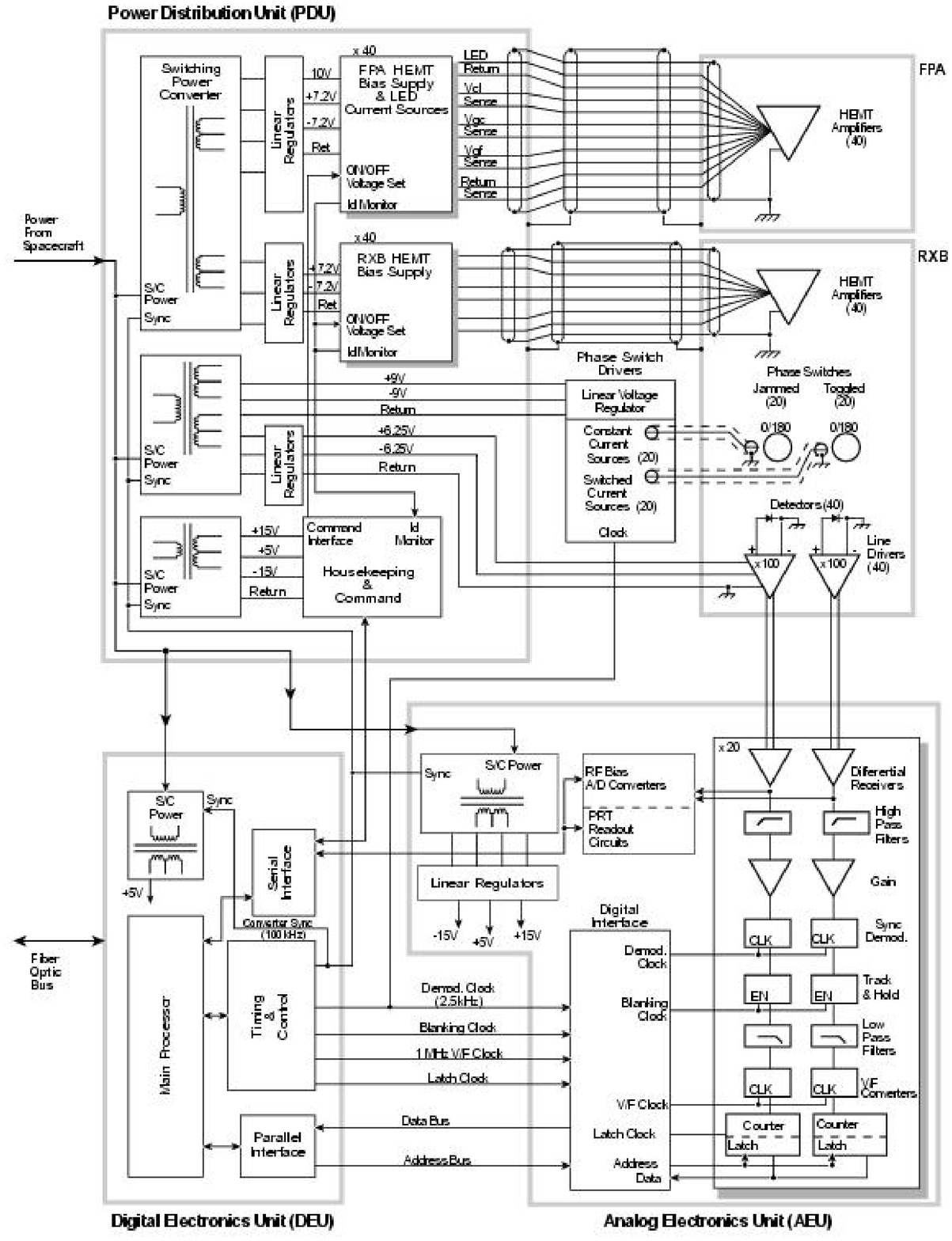}
\caption{Block diagram of the {\sl MAP} instrument  electronics. The PDU, AEU and DEU boxes are located on the main structure
of the spacecraft, and are shaded from direct solar illumination  by the shielding provided by the solar arrays. The RXB 
is located near the 
center of the spacecraft to provide a stable thermal environment. Microwave
 interconnections of the radiometer
components are shown in Figure~\ref{fig:full_rad_dia}.  }
\label{fig:full_inst_dia}
\end{figure*}

\subsection{Nomenclature}
Figure~\ref{fig:full_rad_dia} is a diagram of the {\sl MAP} radiometers.
 The Focal Plane Assembly (FPA) consists of the feed horns, cold portions of the radiometers, and 
structure, all of which operate near 90~K in flight. The Receiver Box (RXB), which comprises the 
warm radiometer components, phase switch driver circuit boards, and structure,  is located near the center 
of the satellite and operates at a nominal temperature of $\approx 285$~K. Microwave signals from the FPA 
are passed to the RXB by 40 thin-wall stainless steel waveguides that provide thermal isolation. 

The decision
to split the radiometer into warm and cold sections in this manner was motivated by performance, scheduling and cost considerations.
Running as few components cold as possible increased reliability by
reducing the number of components undergoing large (100~K $-$ 300~K)
thermal cycles, saved time by allowing most of the component characterization tests to be performed near room temperature, and
eliminated the need for development of cryogenic detectors and preamplifier circuits. The use of warm  HEMT amplification
stages  has performance advantages since $f_\mathrm{amp}$ of HEMT amplifiers operating warm is significantly lower
than for amplifiers operating cryogenically.
This configuration also allowed for an RF tight enclosure to be built around the RXB components, preventing the high level RF 
signals present in these components from 
coupling back to the low signal level FPA section, reducing the possibility of  ring-type oscillations or other artifacts
associated with such feedback. The interior of the RXB and FPA enclosures are
 lined with microwave absorbing
fabric~\footnote{Milliken Research Corp., Contex fabric} to dampen cavity modes present in the enclosure. 
 
Radiometers are identified by a three part designator.
The first part consists of one or two letters (K, Ka, Q, V or W) that specify the nominal 
operating frequency band of the radiometer. The second part consists of a single
digit $(1 - 4)$ that indicates which pair of focal plane feed horns are associated with the radiometer. The
final part also consists of a single digit (1 or 2) and is used to denote which of the
two linear polarizations of the designated feed horn pair the radiometer senses. Designations of `$1$' and `$2$' indicate
that the radiometers are connected to  the main-arm and side-arm of the orthomode transducers (OMTs) respectively. 
  See \citet{opticspaper} for details of the focal plane geometry and
 polarization orientations of each OMT with respect to the satellite. A pair of
radiometers associated with the two polarizations of a given feed  is termed a
Differencing Assembly (DA) and is specified using only the first two elements of the designator, such as
Ka1. 

A diagram of the {\sl MAP} instrument electronics is given in Figure~\ref{fig:full_inst_dia}.
 The Analog Electronics Unit 
(AEU) contains the the data collection circuitry for the radiometers, and the precision readout circuitry
for the platinum resistance thermometers. These thermometers monitor the 
temperature of numerous radiometer components to assist in determining levels of possible
systematic errors.
The Power Distribution Unit (PDU) contains the DC-DC converters that convert the 31.5~V nominal spacecraft
  bus voltage to that required by the phase switch drivers, line drivers,
 and the commandable power supplies which power the FPA and RXB HEMT amplifiers.
 The Digital Electronics Unit (DEU) contains the digital multiplexers and control circuitry
needed to accumulate the science and housekeeping data and pass it to the main
 computer processor over the spacecraft 1773
fiber optic  data bus.
 The AEU, PDU and DEU are mounted on the outside of the hex hub of the
spacecraft structure, are continuously shielded from the Sun by the solar array  and are cooled
 by direct radiation to space.
 More detailed
 descriptions of relevant parts these systems are presented later and in \citet{missionpaper}.
\subsection{ Basic Design }
\begin{figure*}
\epsscale{1.8}
\plotone{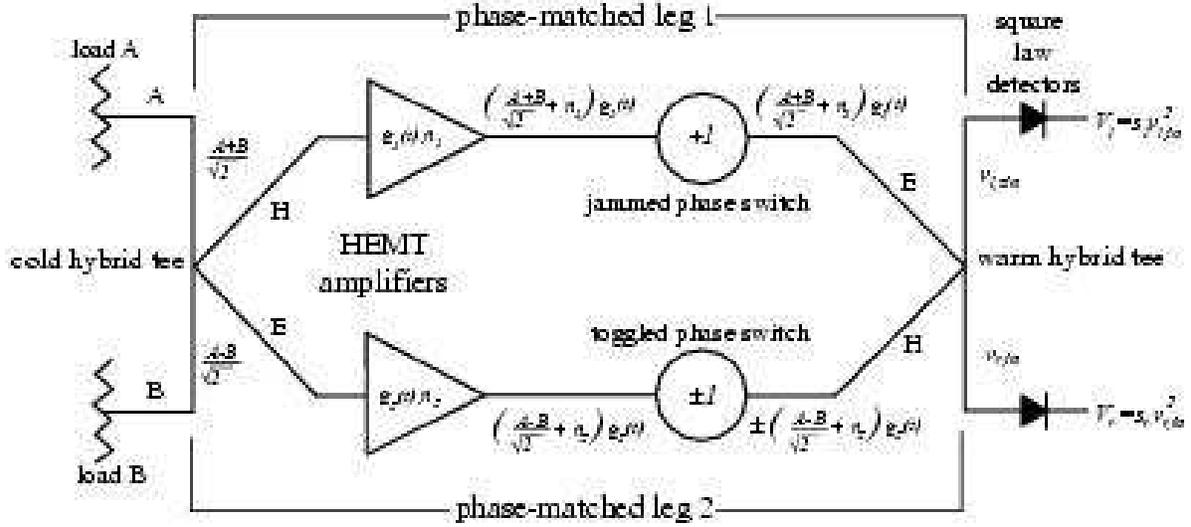}
\caption{Simplified radiometer diagram used in the description of radiometer operation. The amplifiers'
voltage gain and additive noise voltages are given by $g_{1,2}(t)$ and 
$n_{1,2}$. The arms of the hybrid tees are labeled  E and H. Expressions for signal levels are for voltages.}
\label{fig:simple_rad_dia}
\end{figure*}

 Figure~\ref{fig:simple_rad_dia} contains a
slightly simplified version of the radiometer diagram along with the relevant
values of signals present in various sections of the radiometer.  `A' and `B' represent the 
instantaneous, time dependent
voltages emitted by cryogenic input loads, Load-A and Load-B, respectively. These
loads  represent the thermal signal from one linear polarization
of the microwave background radiation.  In the satellite, these signals
are obtained from the output of the
OMT attached to the back of the  feed horns.

For simplicity it is first assumed that most of the components in the 
radiometer are ideal, apart from the amplifiers which are described by
instantaneous input referenced noise \emph{voltages}, $n_i$, and instantaneous
\emph{voltage} gains, $g_i(t)$, where the index $i = 1, 2$ specifies 
in which  phase-matched leg the  components resides. The time dependent
gain of the HEMT amplifiers has been shown explicitly to
emphasize that it is not constant, as is usually assumed.
It is assumed that all the components between the cold hybrid tees and warm hybrid tees are
perfectly phase-matched,  so only terms involving
intentional differential phase shifts between the legs are presented.
The effects of the various non-idealities are discussed in subsequent sections.

Since the signals `A' and `B' are thermally induced noise voltages from two
separate loads they are uncorrelated, and hence obey the relations
\begin{eqnarray}
\overline{A} & = & \overline{B} = 0 \label{eq:input_prop1} \\
\overline{A  B} & =& 0 \label{eq:input_prop2} \\
\overline{A  A} & \propto &  k_B\Delta\nu T_a \\ 
\overline{B  B} & \propto & k_B\Delta\nu T_b 
\end{eqnarray}
where $k_B$ is Boltzmann's constant, $\Delta\nu$ is the bandwidth over 
which the voltages A and B are
measured and $T_a$ and $T_b$ are the physical temperature of the loads
in Kelvin. The over-bar indicates a time average with a period long
compared to the period of the microwave signal, but short compared to
the time scale of instabilities in any of the microwave components.
Given these assumptions, the voltages present at the inputs of the two HEMT
amplifiers are 
\begin{equation}
\frac{A+B}{\sqrt{2}} \ \textrm{and}\  \frac{A-B}{\sqrt{2}}
\end{equation}
where the difference in sign of the B signal reflects the $180^\circ$ relative
phase shift between the signals in the phase-matched leg introduced
by the cold hybrid tee. The signals after amplification by the HEMT amplifiers  become
\begin{eqnarray}
u_1 = \left(\frac{A+B}{\sqrt{2}} + n_1\right)g_1(t) \label{eqn:u_def1}\\
u_2 = \left(\frac{A-B}{\sqrt{2}} + n_2\right)g_2(t) \label{eqn:u_def2}
\end{eqnarray}
where $n_1$ and $n_2$ are the input referenced noise voltages added by
the cryogenic HEMT amplifiers.

The phase switches are used to introduce  additional 0 or 180$^\circ$
phase shifts between the phase-matched legs. To accomplish this one of them
is left `jammed' in one state (here chosen as leg 1) to provide 
the correct insertion phase
while the other (leg 2) is switched between two states. The signals
 at the output of the
warm hybrid tee are then
\begin{eqnarray} 
v_{l,\mathrm{in}} & = & \frac{1}{\sqrt{2}} \left(\frac{A+B}{\sqrt{2}}
    +n_1\right)g_1(t) \nonumber\\
    &  & \mbox{} \pm \frac{1}{\sqrt{2}}\left(\frac{A-B}{\sqrt{2}}
    +n_2\right)g_2(t) \label{eq:warm_hybrid_output_left}
\end{eqnarray}
and
\begin{eqnarray} 
v_{r, \mathrm{in}} & = & \frac{1}{\sqrt{2}} \left(\frac{A+B}{\sqrt{2}}
    +n_1\right)g_1(t) \nonumber\\
    & & \mbox{} \mp \frac{1}{\sqrt{2}}\left(\frac{A-B}{\sqrt{2}}
    +n_2\right)g_2(t)
\end{eqnarray}
where the upper sign refers to the signal with the phase switch in the
$0^\circ$ relative phase state and the lower sign to the switch in the 
$180^\circ$ state. The detectors are operated in the square law region
where their output voltage is proportional to the square of their input
voltage. The response of these are represented  as
\begin{equation}
V_l = s_l v_{l,\mathrm{in}}^2\ \textrm{and}\ V_r = s_r v_{r,\mathrm{in}}^2
\end{equation}
for the left and
right detectors where $s_l$ and $s_r$ are the responsivities
of the two detectors. In general these responsivities are a function of the 
input microwave frequency, but for now  will be assumed   
frequency independent and equal so that $s_l = s_r = s$.
 The outputs of the two detectors
then become
\begin{eqnarray}
V_l & = & \frac{s}{2}\ \biggl\{ \left(\frac{A^2+B^2}{2} 
            + n_1^2\right)g_1^2(t) \nonumber \\
    &   & \mbox{} + \left(\frac{A^2+B^2}{2} 
            + n_2^2\right)g_2^2(t) \nonumber \\
    &   & \strut \mp (A^2-B^2)g_1(t)g_2(t)\biggr\}
\end{eqnarray}
and
\begin{eqnarray}
V_r & = & \frac{s}{2}\ \biggl\{ \left((\frac{A^2+B^2}{2} 
            + n_1^2\right)g_1^2(t) \nonumber \\
    &   & \mbox{} + \left(\frac{A^2+B^2}{2} 
            + n_2^2\right)g_2^2(t)  \nonumber \\
    &   & \pm (A^2-B^2)g_1(t)g_2(t)\biggr\}
\end{eqnarray}
where equations \ref{eq:input_prop1} and \ref{eq:input_prop2} have been used together with the  fact that the intrinsic voltage
noise of the HEMT amplifiers is uncorrelated to the signals from either
of the loads, i.e.
\begin{equation}
\overline{ n_i  A} = \overline{n_i  B} = 0.
\end{equation}
The outputs consist of three terms. 
The first two terms arise from the sum of the mean power of the radiometer inputs signals, $\frac{A^2+B^2}{2}$, and the
noise power added by the cryogenic HEMT amplifiers, $n_i^2$. The contribution arising from the radiometer input signal is
proportional to the antenna temperature of the input signals, while the term from the HEMT amplifier's noise is proportional
to the HEMT amplifier's input-referenced noise temperature. These amplifier noise temperatures  range
 from 30~K $-$ 96~K for K-band through
W-band respectively. Both these terms are multiplied by the time dependent
gain terms $g_i^2(t)$, and as such fluctuate due to small gain variations of
the amplifiers. Because of the large system bandwidth, 
even small gain variations can substantially impact the performance of the
radiometers. For example, at W-band the nominal effective radiometric bandwidth is $\Delta \nu_\mathrm{eff} =19$~\GHz,
so the fractional change in gain that produces a signal equal to the 
intrinsic
radiometric noise after integrating for one second is,
\begin{equation}	
\frac {\Delta g}{g} = 1 / \sqrt{\Delta \nu  \tau} = 7.3 \times 10^{-6}.
\end{equation}

 The signal
of interest is the third
term, which is the difference between the antenna temperatures of
 the two radiometer
inputs, and is multiplied by the product of the voltage gains of the
 two amplifiers.
The magnitude of the prefactor ${(A^2 - B^2)}$ is much smaller 
that that
of the first two terms, typically in the mK regime. 

There are two techniques
 for
recovering this signal in the presence of the signals from the first two terms,
which dominate the detector outputs.
 The first is to notice that the first two terms
appear identically on the output of both detectors
 whereas the signal of interest
appears with the opposite sign on the two detectors.
 If the output of the detectors
are differenced,  the offending terms cancel and the desired terms add. This
assumes that the responsivities of the two detectors are identical. The other
 method
is to difference the output of each detector with the phase switch in
 opposite states.
In this case the first two terms remain constant and hence cancel, while
 the third
term changes sign when the phase switch flips, and therefore produces
 a net output.
For this method to work the time scale at which the phase switch is switched and the
differencing occurs must be short compared to the time scale of the instabilities
in the radiometer components,
otherwise the value of the first two terms will change in the interim and therefore
will not cancel. 

The {\sl MAP} radiometers utilize both techniques. The phase switches
are toggled at 2.5~\kHz\ and on board lock-in amplifiers perform the synchronous
differencing of each detector's output. The output of the lock-in associated with
each detector's output is then averaged and telemetered to the ground where the differencing
between pairs of detectors is performed. This combination provides optimal rejection
of various instabilities. For example, if only the detector - detector differencing
were performed low frequency drifts in the detector responsivity and offset voltage
drifts of the following video amplifiers would limit the performance of the radiometer.
On the other hand if only the temporal differencing were performed the switching
frequency of the phase switches would have had to be substantially increased due
to the proximity of the $1/f$ knee of the W-band amplifiers. This would
have increased the `dead time' of the detectors due to the blanking interval and hence reduced
the sensitivity of the radiometers. (See Section~\ref{sec:aeu}.)

\begin{figure*}
\epsscale{1.2}
\plotone{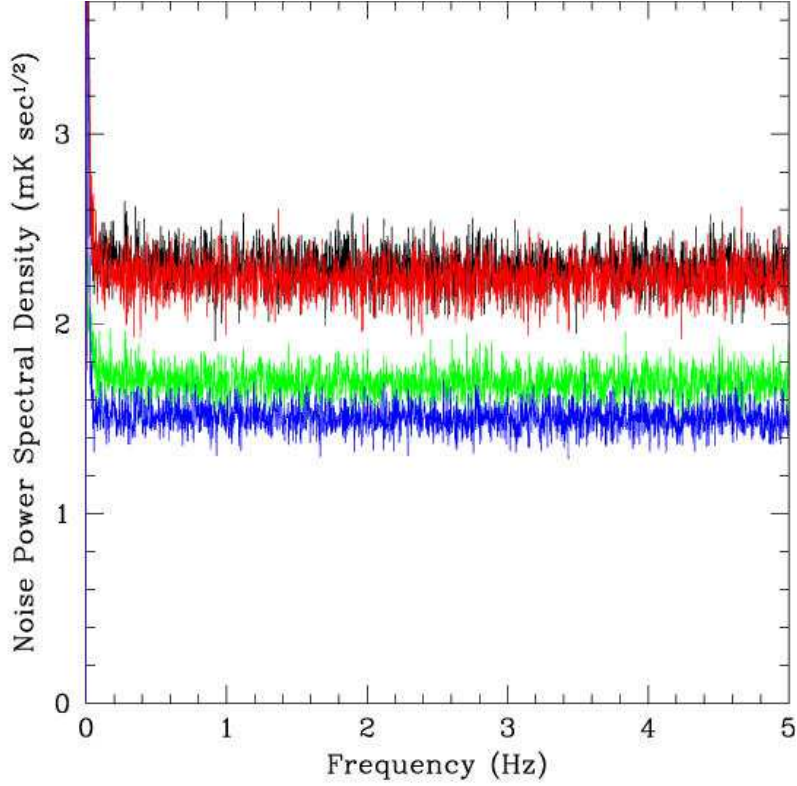}
\caption{Power Spectral Density of the {\sl MAP} W11 radiometer. The red and black traces are the power spectral density measurements of the 
two detectors on the radiometer. The green and blue traces are the spectral 
densities of the normalized sum and difference
of the detector signals. During these measurements the FPA HEMT amplifiers were operating
 at a physical temperature of
$\approx 75$~K and the loads attached to the inputs of the radiometer were regulated to 25~K. }  
\label{fig:W11_011PSD}
\end{figure*}

Figure~\ref{fig:W11_011PSD} presents a noise power spectral density measurement of the W11 radiometer made
during the radiometer assembly and testing program. Demodulated data for each detector, L and R, 
and the normalized mean, $(\mathrm{L}+\mathrm{R})/2$ and 
difference $(\mathrm{L}-\mathrm{R})/2$ of the detector 
signals are displayed. The noise
 power spectral densities are flat
to very low frequencies as expected, despite the fact that the amplifiers' $f_\mathrm{knee}$ frequency for
 the RF bandwidth
of this radiometer is $\approx 1$~kHz. In the absence of gain fluctuations, the difference signal between the
detectors should have a noise power level lower by a factor of $\sqrt{2}$  than each detector's individual
noise. Note that the noise power spectral density of the difference signals is
 lower that the individual detector
noises by \emph{more} than this factor. This occurs because there is still a substantial degree of amplifier
gain fluctuations present at the 2.5~\kHz\ phase switch drive frequency. This contributes additional correlated
noise
 to both detectors, in addition  to the
intrinsic radiometer noise. When the detector signals are differenced this additional correlated term 
cancels, as described above, resulting in a noise power reduction of greater than $\sqrt{2}$ relative to the individual detector noises.

\subsection{Departures From Ideality}
Non-idealities in the components comprising the radiometers lead to effects not described by the
previous analysis. 
A detailed model simultaneously
incorporating all such effects would prove complicated and
not readily understandable. Instead, the largest and most important
effects are described individually so as to provide an understanding of
how each term relates to the overall radiometer performance. In doing so, 
it is important to realize that  the radiometers' `signal' and
`noise' are often altered differently by these effects.

\subsubsection{Chain to Chain Amplitude and Phase Match} 
Since the switching  is accomplished through phase
modulation, it is important to understand how phase and amplitude mismatches
between the two phase-matched legs affect radiometer performance. There are two types of phase
 error that can occur.
Nominally there should be a $180^\circ$  difference in the insertion phase of the phase switch between its
two states.  Fortunately the phase switches are excellent in this respect, with departures
 from ideality of  $< 3^\circ$. Such
small errors do not significantly affect the radiometers' performance. The more important
phase error is the departure from the $0^\circ$  (or $180^\circ$) phase match between the
 two legs of the radiometer.
These errors are typically much larger since each leg comprises several components, each of which
can contribute to the mismatch. 

\begin{figure*}
\plotone{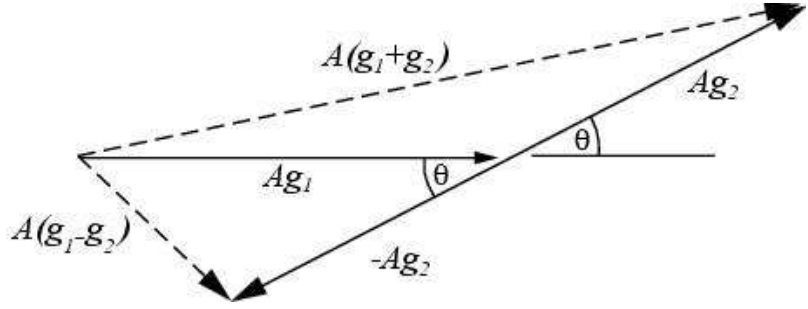}
\caption{Vector diagram showing the relative phase angle of the amplified voltage signals
from source A for both phase switch states. The relative phase shift between the phase-matched legs
is  $\theta$.}
\label{fig:phase_err}
\end{figure*}

First consider effects  on the signal terms. Equation~\ref{eq:warm_hybrid_output_left}
 describes the input voltage to 
one of the square law detectors.
Ideally $ g_1(t) = g_2(t)  = g $, so signals originating from source A travel through both legs
 of the radiometer and appear at the input of the square law detector with equal amplitude
and the same (or opposite) phase so that that they add (or cancel).
 The signal at the output of the square law detector
from source A therefore is either  $\frac{s}{2} (A^2g^2) $ or $ 0 $, depending on the phase switch state.
 In the presence of 
phase and amplitude mismatches the voltage signals at  the input of the detectors for the two
phase switch states  become
$\frac{1}{2}A (g_1(t) + g_2(t)) $ and $\frac{1}{2}A (g_1(t) - g_2(t))$. After square law detection  and 
 demodulation the
resulting signal is $ \frac{s}{2}A^2g_1(t)g_2(t) \cos(\theta)$, where $\theta$ is the phase error as shown
 in Figure~\ref{fig:phase_err}.
Signals from the `B' source are affected similarly, so the net effect is that the amplitudes of the 
modulated signals
output from each detector, $S_i$, are
\begin{eqnarray}
S_l \propto - s_l (A^2 - B^2)g_1(t)g_2(t) 
    \cos(\theta)  \label{eq:det_signal_l}\\
S_r \propto + s_r (A^2 - B^2)g_1(t)g_2(t) 
    \cos(\theta)   \label{eq:det_signal_r}.
\end{eqnarray}

The noise is affected quite differently. The \emph{noise}  in the two phase-matched legs of the radiometer
is uncorrelated since it originates from different amplifiers. The noise voltages present at the
 input of the left and right detectors
 are $ \frac{u_1 + u_2 }{\sqrt{2}}$ and $ \frac{u_1 - u_2 }{\sqrt{2}}$, where $u_1$ and $u_2$ are  defined by
 Equations~\ref{eqn:u_def1} and ~\ref{eqn:u_def2}. The corresponding instantaneous input powers 
to the detectors are $ \frac{(u_1 + u_2)^2}{2}$ and $ \frac{(u_1 - u_2)^2}{2}$, while average power
 on both detectors is the same and is given by
$ \frac{\overline{u_1^2} + \overline{u_2^2}}{2}$.
 The detectors operate in the square law region, so the instantaneous and mean detector voltages are
\begin{eqnarray}
V_l =  \frac{s_l}{2}(u_1 + u_2)^2, 
& \overline{V_l} = \frac{s_l}{2}( \overline{u_1^2} + \overline{u_2^2})\\
V_r =  \frac{s_r}{2}(u_1 - u_2)^2, 
& \overline{V_r} = \frac{s_r}{2}( \overline{u_1^2} + \overline{u_2^2})  .
\end{eqnarray}
The variances of the detector output voltages, 
$N_i^2$ --- measures of the radiometer noise --- are proportional to the
square of the total power incident on each detector  and are given by
\begin{eqnarray} \label{eq:single_det_noise}
N_l^2 = \overline{V_l^2 - \overline{V_l}^2} 
= \overline{V_l^2} - \overline{V_l}^2 
= \frac{s_l^2}{2}(\overline{u_1^2} 
+ \overline{u_2^2})^2  \label{eq:det_noise_l} \\
N_r^2 = \overline{V_r^2 - \overline{V_r}^2} 
= \overline{V_r^2} - \overline{V_r}^2 
= \frac{s_r^2}{2}(\overline{u_1^2} 
+ \overline{u_2^2})^2. \label{eq:det_noise_r} 
\end{eqnarray}
 The variance of the \emph{difference} between the detector voltages, $N_\mathrm{diff}^2$,  is
\begin{eqnarray}
N_\mathrm{diff}^2 &=& \overline{(V_l-V_r)^2}-\overline{(V_l-V_r)}^2 \\
&=&(\overline{V_l^2}-\overline{V_l}^2) + (\overline{V_r^2}-\overline{V_r}^2) \nonumber \\
& &\mbox{}- 2\overline{(V_l-\overline{V_l})(V_r-\overline{V_r})} \\
&=& s^2(\overline{u_1^2} + \overline{u_2^2})^2 - s^2 (\overline{u_1^2}-\overline{u_2^2})^2 \\
&=& 4  s^2 \overline{u_1^2} \ \overline{u_2^2} \label{eq:det_noise_diff}
\end{eqnarray}
where the relation
\begin{equation} \label{eqn:det_cov}
\overline{(V_l-\overline{V_l})(V_r-\overline{V_r})} = \frac{ s^2}{2}(\overline{u_1^2}-\overline{u_2^2})^2
\end{equation}
has been used to relate the covariance of the detector output voltages to the power
 incident to the two inputs of the
warm hybrid tee as shown in  Appendix~\ref{app:noise_cov}. It has also been assumed that the responsivity of the two detectors are identical, 
$s_l = s_r = s $. This is achieved in practice by separately telemetering the signals from the
two detectors to the ground and individually calibrating them before calculating the difference.

It is a good approximation to assume that the input referenced noise temperature of the two amplification
chains within a radiometer are equal, $n_1 = n_2$.
The signal to noise ratio, $\frac{S_i}{N_i}$, for \emph{each} detector then depends on  the phase and gain
 mismatch between the phase-matched legs, (Equations \ref{eq:det_signal_l}, \ref{eq:det_signal_r}, \ref{eq:det_noise_l}, \ref{eq:det_noise_r})
\begin{equation}
\frac{S_i}{N_i} \propto \frac{\overline{g_1} \ \overline{g_2}\cos(\theta)}{\sqrt{\overline{g_1^2} + \overline{g_2^2}}},
\end{equation}
however, the signal to noise ratio for the \emph{difference} of the detector outputs is \emph{independent} of the
gain  mismatch, (Equations~\ref{eq:det_signal_l}, \ref{eq:det_signal_r}, \ref{eq:det_noise_diff} )
\begin{equation}
\frac{S_r - S_l}{N_\mathrm{diff}} \propto \cos(\theta).
\end{equation} 
This behavior is similar to that of a true correlation radiometer, and results from the existence of the
detector noise covariance as described in Appendix~\ref{app:noise_cov}.

The covariance between the output of the two detectors (Equation~\ref{eqn:det_cov}) of
each radiometer was used as a diagnostic to verify the 
health of the radiometers during integration and testing. Any change from the nominal value measured during
radiometer assembly would indicate a change in the relative RF gain of the amplification chains. Similar data
are available during flight, since the two detector outputs are separately telemetered to the ground and provides
a continuous monitor of the gain balance between the two phase-matched legs of each radiometer.

The {\sl MAP} design  specification was to hold the gain imbalance between chains to $\pm 2$~dB and 
the phase difference to $\pm 20^\circ$, making the loss in sensitivity from the phase error $<$  6\%.
 It should be noted that although the above
relation describes the independence of the signal to noise ratio from amplifier gain imbalances, any
other sources of gain mismatch, due to imbalances in the hybrid tees or losses in the waveguides
or phase switches affect the signal to noise ratio in the exact same way, so the gains in the previous
equations  can be thought of as the gain of the entire system between the hybrid tees.

\subsubsection{Crosstalk in the Cold Hybrid Tee}
Imperfect isolation between the two ports of the cold hybrid tee connected to the
cryogenic HEMT amplifiers is a potential source of a radiometric offset. Consider
the noise voltage, $n_1$, emitted from the input of one of the cryogenic HEMT amplifiers.
The magnitude of this voltage will roughly correspond to the input referenced noise
 temperature of the HEMT amplifier. This noise will
have some correlation with the output voltage of the HEMT amplifier \citep{wedge}.
 Let the correlation coefficient between these voltages
be $\beta$. Although the value of this coefficient is not well known, it is expected to
fall in the range of 0.1 to 0.3. If the \emph{power} isolation between the E and H  ports of the hybrid tee
is $\alpha $, the resulting voltage at the input of the other cryogenic HEMT amplifier produced
by the noise voltage output from the first will be
\begin{equation}
	v_\mathrm{in2} = n_1 \sqrt{\alpha}
\end{equation}
where typical values of $\alpha$ are $\approx 1 \times 10^{-4}.$
Since this signal is coherent between the two phase-matched legs of the radiometer, it can
produce an additional contribution to the total detector power that is modulated by the
phase switch. The size of this effect will be roughly
\begin{equation}
	T_\mathrm{off} \approx (n_1^2 + n_2^2)  \beta \gamma \sqrt{\alpha}
\end{equation}
where $\gamma$ is an additional  factor that allows for the effect
of bandwidth averaging. Its magnitude is
roughly $\gamma \approx l_c / \Delta l $ for $ \Delta l > l_c$, where $l_c \approx c / \Delta \nu_\mathrm{eff}$ is
 the coherence length of the microwave signal,
determined by the bandwidth of the radiometer, $\Delta \nu_\mathrm{eff}$, and $\Delta l \approx 5\cm$ is the path length difference
between the two paths the signals travel before being recombined in the warm hybrid tee. Since
noise emitted from both amplifiers contributes to this effect, its magnitude is proportional to the sum
of the noise temperatures. The desire to minimize the size of this effect led to the configuration in which the 
amplifiers are attached to the E and H arms of the hybrid tees, since the E-H arm isolation is
 typically 10~dB better
than the colinear-colinear arm isolation.

	There is an additional term that can also contribute to this effect caused by reflections
 from either the OMT or the corrugated feed. In this case the power emitted from
the input of the HEMT amplifier enters the cold hybrid tee through an E or H-port and is coupled to the colinear
ports. This signal is then reflected from the OMT or feed horn, re-enters the cold hybrid through the colinear
ports and is coupled to the opposite  H or E-port. This effectively reduces the E-port to H-port isolation of the
cold hybrid tee. The contribution from this effect to a radiometer's offset is $\approx 0.2$~K.

\subsubsection{ Toggled Phase Switch Transmission Imbalance }
Slight differences between the frequency dependence of the toggled phase switch's transmission coefficients 
 in its two states can lead to an effective radiometer offset.
 This occurs if the frequency responses of the 
two detector/filter combinations  on a radiometer differ. The detector
 output voltage for detector $j$ with the phase switch in state $i$
can be expressed as
\begin{equation}
	V_{i, j} = \int u^2(\nu)t_i(\nu) s_j(\nu)d\nu + V_j^0
\end{equation}
where $u^2(\nu)$ is the RF noise power spectral density entering the toggled phase
 switch, $t_i(\nu)$ is the frequency
dependent power transmission coefficient of the phase switch in the two
 states, $i = 1, 2$, and $s_j(\nu)$ is the
frequency dependent responsivity of the detector filter combination, $j = L, R$. The integral
 extends over the RF bandwidth of the 
radiometer, and $V_j^0$ is a constant detector output voltage arising from
 the noise power from the unswitched leg
 of the radiometer.

 Consider the situation where all other sources of
 radiometric offset vanish. The condition to null the \emph{modulated}
component of the right detector output voltage is $V_{1, R} = V_{2, R}$, however this does not ensure that 
the modulated component of the left detector's output voltage will vanish, $V_{1, L} = V_{2, L}$, unless
$s_L(\nu) = s_R(\nu)$. In fact, it raises the  question
 as to the definition of the offset for this type of radiometer.
The definition used by {\sl MAP} is that a radiometer's offset is the temperature difference
 between thermal loads connected to the
 radiometer's inputs that produces the 
condition $V_{1, R} - V_{2,R} = V_{1, L} - V_{2, L}$.
 Once this condition is achieved by setting the temperature of the test input loads,
 the difference between the average losses of the two states of
 toggled phase switch is adjusted to make both $V_{1, R} - V_{2,R} = V_{1, L} - V_{2, L} = 0$ by
 adjusting the phase switch bias currents.
 This definition of offset has the advantage that it is independent of the sign and gain
of the data collection system used to record the detector output voltages.

\subsubsection{ Standing Waves} \label{sec:standing_waves}
Standing waves between the cold and warm HEMT amplifiers degrade the phase 
match and gain flatness of the radiometers, with the
associated reductions in sensitivity previously described. Traditionally, isolators
 are used to eliminate standing waves,
however the only
viable isolator for {\sl MAP} radiometers, given the large fractional bandwidths, would 
be Faraday rotation isolators. Such isolators
unfortunately exhibit small changes in characteristics under the influence of externally applied
 magnetic fields. The Cosmic Background
Explorer (COBE)
used isolators \citep{cobepaper}, as well as ferrite switches, and needed to correct the data for variations resulting from the 
changing orientation of Earth's magnetic fields with respect to the spacecraft, even though efforts were made to 
magnetically shield the ferrite devices. Despite the fact that {\sl MAP} operates far from Earth, and therefore
in a much smaller magnetic field, $\approx 100$~nT, there were still concerns related
 to magnetic fields generated on board the spacecraft from
 the reaction wheels and other high current spacecraft systems. 

For the higher frequency bands the loss in the phase-matched waveguides is sufficient to reduce the standing waves to
 acceptable levels. For example, at W-band the loss in the 
phase-matched waveguides
 is approximately 6~dB, so the largest
amplitude and phase variations resulting from standing waves are approximately $\pm 2.2$ dB and $\pm 14 ^\circ$
respectively, even if the inputs of the warm amplifiers and outputs of the cold amplifiers had 0~dB return losses.
With typical input and output return losses of 4~dB  and 10~dB these variations in amplitude and phase
are approximately $\pm 0.4$~dB and $\pm 3 ^\circ$. 
At lower frequencies the loss in the interconnecting waveguide is much smaller, typically 1~dB. When combined with
the typical input and output return losses of the amplifiers of about 8~dB, the amplitude and phase variations
are approximately $\pm 1.2$~dB and $\pm 8^\circ$. For all frequency bands the performance degradation
resulting from these effect was small enough that no isolators were used.

Standing waves between the outputs of the RXB amplifiers and the detectors are also of concern.
 Not only do they
introduce gain flatness errors and phase errors  as previously described, but,
 since the toggling phase switch modulates
these standing waves, they can introduce an effective offset into the
radiometer if the frequency responses of the two filter/detector combinations 
on a radiometer are not identical. This effect is similar to that
resulting from the transmission
mismatch of the phase switch, but in this case
the spectrum of the available noise power, determined by the standing wave mode pattern, is modulated by
the state of the phase switch.
 This effect was minimized by using detectors with input matching circuits
that reduced their reflection coefficients.

\section{ Systematic Errors }
  Potential sources of radiometer induced signals contributing to the systematic error budget included random
 output drifts as well
as terms arising from temperature and power supply fluctuations, varying magnetic fields, either external or internal to the spacecraft,
microphonics, and particle hits. 
The {\sl MAP} radiometer systematic error analysis categorized errors into two types,
additive and multiplicative. Additive signals are those that add linearly to the radiometer output signals, such as varying emission from
the optics or feeds. Multiplicative signals are those that modulate the effective gain of the radiometer and therefore require a non-zero
radiometer output to have affect. In practice the only significant multiplicative terms  are those arising from the
product of small radiometer gain fluctuations and the radiometers'  offsets. For the purpose of estimating the size of these terms,
offsets were assumed to be 1~K for all
radiometers. 
 Spin synchronous effects arise when some driving function coupled to the spacecraft orientation, such as a temperature
 or power supply voltage, induces a change in the output of a radiometer by varying the characteristics of one or more of 
 its components. 

During development the magnitudes of the estimated systematic errors were calculated based on current estimates of the 
maximum size of the
expected driving functions for each component. The driving functions were  multiplied by susceptibility coefficients relating how
each ultimately perturbed the output of the radiometers through either  additive or multiplicative terms. 
The susceptibility
coefficients were derived by measuring the variation of important component parameters as a function of the
 driving parameters. It was often necessary to vary the driving parameter over a much larger range than expected
in flight in order to measure a susceptibility coefficient. In such cases it was assumed that the component's 
response scaled linearly
with the driving function.
The  radiometer model was used to calculate the resulting perturbations of the radiometer's outputs. 
  Given the symmetric design of the observatory, and the long thermal time constants relative to the 
spin period of the spacecraft, it was assumed that the various systematic error terms have random phases relative to one another, so
systematic error contributions were added in quadrature. For the purpose of design, the spin synchronous temperature variation of the FPA and
 RXB components was  assumed to be 0.5~mK peak, and the spin-synchronous temperature fluctuations
 of the AEU, DEU and PDU components was assumed
to be 10~mK peak. Table~\ref{table:sys_err_terms} 
summarizes the radiometer related systematic
error terms identified. This analysis was used to evaluate trade 
offs and tests used throughout
the development process, but does not necessarily 
reflect the amplitude of the effects in flight data. 

\begin{table*}[t]
\caption{\small  Systematic Error Terms  }
\small{
\vbox{
\tabskip 1em plus 2em minus .5em
\halign to \hsize {#\hfil & #\hfil &#\hfil \cr
\noalign{\smallskip\hrule\smallskip\hrule\smallskip}
Component     & Susceptibility $\times$ Forcing function  & Effect \cr
\noalign{\smallskip\hrule\smallskip}
Feed horns              & $\epsilon \times \Delta (T_a - T_b )$         & A \cr
Orthomode transducer    & $\epsilon \times \Delta (T_a - T_b )$    & A\cr
OMT - hybrid waveguides & $\epsilon \times \Delta (T_a - T_b )$      & A\cr 
FPA hybrid tee          & $ (\epsilon_a - \epsilon_b) \times \Delta T$      
& A\cr
FPA HEMT amplifiers     & $dg/dx \times \{\Delta V_\mathrm{gc}, 
\Delta V_\mathrm{gf}, \Delta V_d, \Delta T, \Delta I_\mathrm{LED}\}$&M\cr
RXB HEMT amplifiers     & $dg/dx \times \{\Delta V_\mathrm{gf}, 
\Delta V_d, \Delta T$ \}                              & M\cr
Band definition filters & $dS_{21}/ dt \times\Delta T$  & $A^*$,M\cr
Phase switches          & $d(S_{21}(0^\circ) - S_{21}(180^\circ) / dx 
     \times \{\Delta T, \Delta I$\}  & $A^*$,M\cr
Detectors               & $ds/dT \times \Delta T$ & M\cr
Line drivers            & $dg/dx \times \{\Delta V_\mathrm{dd}, 
\Delta V_\mathrm{ss}, \Delta T$ \}  & M\cr 
AEU & $\{d A_v/d x, d O_v/ d x\} \times \{\Delta T, 
\Delta V_\mathrm{bus} \}$ & A, M\cr
PDU & $\{d V_\mathrm{gc}/d x, d V_\mathrm{gf}/d x, d V_d/d x, 
d I_\mathrm{LED}/d x \} \times \{ \Delta T, \Delta V_\mathrm{bus}$ \} & \cr
\noalign{\smallskip\hrule}
}}}
\small{
 The forms of systematic error terms incorporated into the model
used to estimate the size of artifacts on the radiometers' outputs. The first column lists
 the component originating the effect. The
second column contains the relation describing the component's varying performance characteristic in 
the form \emph{coefficient} $\times$ \emph{forcing function}.
The \emph{coefficient} is either a fixed parameter such as an emissivity, $\epsilon$, or
 the derivative of a parameter, $d/dx$,
with respect to a forcing function $x$. For the AEU $A$ and $O$ represent the gain and offset of the 
entire data collection system from the line driver inputs  to the digitized outputs. For the PDU and HEMT amplifiers
$V_\mathrm{gc}$, $V_\mathrm{gf}$, and
$V_d$ refer to the gate and drain power supply voltages  and $I_\mathrm{LED}$  the LED current supplies as described in Sections
 ~\ref{sec:hemt_amp} and \ref{sec:hemt_bias_supply}.  For the filters and phase switches $S_{21}$ is the forward
transmission element of the scattering matrix.  The \emph{forcing function }
describes which external variable  changes to induce the effect. Temperatures of the component in the first column 
are indicated by $T$. For the phase switches $I$ refers to the PIN diode drive current. The line driver
power supply voltages are $V_\mathrm{dd}$ and $V_\mathrm{s}$, and $V_\mathrm{bus}$ refers to the voltage of main spacecraft power.
Subscripts `$a$' and `$b$' refer to items associated with the different focal planes.
 When more than one parameter or forcing function is
 important they are enclosed
 by braces. The final column
 indicates the form of the perturbing term, `M' for
multiplicative terms and `A' for additive terms. Terms marked $A^*$ are additive terms
 that effect both detectors of a radiometer
in common mode, and so should cancel when the detector outputs are differenced as the
 data is processed. It was estimated that 10~\% of
the effect could survive the differencing, so that was the proportion of the effect
 included in the error analysis.
Variation in the PDU's performance produces driving functions that are used as inputs to other terms.
}
\label{table:sys_err_terms}
\end{table*}

During radiometer assembly and testing, 
the overall radiometer susceptibility to 
varying temperature, power supply voltage,
and magnetic fields was measured to ensure that the actual radiometer susceptibilities met specification.
These susceptibility coefficients were remeasured after the 10 DAs were integrated into the instrument structure in a series of
tests performed in thermal-vacuum chambers at Goddard Space Flight Center (GSFC). Tests were performed both before
 and after an instrument level vibration,
 and results were compared to ensure that no significant changes in the radiometers' performance occurred as a result of the vibrations. 
 During these tests the radiometers were
operated by the flight electronics and in an environment close to that expected in flight. The radiometers' inputs were
 attached to the flight feed horns, each of which
had a full aperture temperature regulated cryogenic load attached to its open end to mimic the microwave background radiation signal.  Following the integration of 
the instrument to the spacecraft
 another set of thermal vacuum  tests was performed in  the Space Environment
 Simulator to search for signals induced in the radiometers from interactions with spacecraft systems.

The data obtained
during all the aforementioned tests were used to calculate approximate susceptibility coefficients. Measurements were also performed during the thermal
vacuum tests to measure the size of the driving functions expected in flight.
 These data, in combination with the in-flight measurements of numerous temperatures,
 voltages and currents in the radiometer system,
 will be used to estimate and/or limit the magnitude of systematic error
signals produced in flight. The results of the instrument level and spacecraft level tests indicate that the {\sl MAP}
radiometers will meet their systematic error requirements in-flight \emph{ without applying any corrections}
 to the radiometric data.

\section{Components}
The {\sl MAP} radiometers are built from numerous discrete components, with little integration of multiple
functions within a component. This approach eliminated the 
design time associated with custom components, allowed for simple characterization and testing of
each component, and for the selection and matching of component characteristics when required.
 It also increased the number of potential suppliers for components since
the number of suppliers of highly integrated components is very limited.

 Most of the microwave components
were of standard commercial quality, with GSFC providing quality assurance oversight, including
materials analysis and qualification testing.  Microwave
joints were wrapped with flexible microwave absorber~\footnote{Emerson \& Cuming Eccosorb BSR-2/SS-6M }
wherever possible to block potential leakage. Important performance characteristics of the
major components of the {\sl MAP} radiometers are provided in the following sections.

\subsection{HEMT Amplifiers}\label{sec:hemt_amp} 
The HEMT amplifiers at the heart of the radiometers were designed, fabricated and tested at the
National Radio Astronomy Observatory (NRAO) Central Development Laboratory
\footnote{ National Radio Astronomy Observatory,
                                      2015 Ivy Road Suite 219,
                                      Charlottesville, VA 22903 } \citep{amplifierpaper}.
 The HEMT devices were 
manufactured by Hughes Research Laboratories\footnote{Hughes
 Research Laboratory, Malibu, CA 90265}.
Key amplifier performance parameters, summarized in Table~\ref{table:hemt_specs},   include noise 
temperature, gain flatness, RF bandwidth and power dissipation. 

\begin{table*}[t]
\caption{\small Flight Radiometer HEMT Amplifier Specifications   }
\small{
\vbox{
\tabskip 1em plus 2em minus .5em
\halign to \hsize { #\hfil &\hfil#\hfil &\hfil#\hfil &\hfil#\hfil
                  &\hfil#\hfil &\hfil#\hfil \cr
\noalign{\smallskip\hrule\smallskip\hrule\smallskip}
Frequency range (\GHz) & 20-25 & 28-37 & 35-46 & 53-69 & 82-106 \cr
Noise temperature @85~K (K) &$30 \pm 2$  &$33 \pm 3$ &$ 48 \pm 5$ &$64 \pm 7$ &$ 96 \pm 7$ \cr
Noise temperature @300~K (K) &$ \approx 100 $  &$ \approx 115  $ &$ \approx 150  $ &$ \approx 260  $ &$ \approx 370 $ \cr
Gain (@85~K/@300~K) (dB) & 34/33 & 34/32 & 34/31 & 35/31 & 35/32 \cr
Gain flatness (dB) & $\pm 1.5$ & $\pm 1.0$ & $\pm 1.5$  
& $\pm 2.5 $ & $\pm 3 $ \cr
\# gain stages & 4 & 4 & 4 & 5 & 6 \cr
Dissipation(RXB) (mW) & $\approx 40$ & $\approx 45 $ 
& $\approx  45$ & $\approx 45 $ & $\approx 55$ \cr
Dissipation(FPA)\tablenotemark{a} (mW) & $\approx 22$  
& $\approx 30 $   & $\approx 25 $  & $\approx 12 $  &  $\approx 25$ \cr
Mass (kg)         &0 .131         & 0.121           & 0.132 & 0.114 & 0.116 \cr
Phase match (degrees) & $\pm 15$ & $\pm 15$ & $\pm 20$ & $\pm 20$ 
& $\pm 25$ \cr
\noalign{\smallskip\hrule}
}}
\tablenotetext{a}{The  values given do not include the $\approx 9$~mW dissipation of the LEDs on each FPA amplifier.}
}
\label{table:hemt_specs}
\end{table*}

The amplifiers are built  using discrete devices, stripline
matching elements and wire bonded interconnections. 
Early in the program NRAO demonstrated the feasibility of
building amplifiers at {\sl MAP}'s highest frequency band 
with highly reproducible characteristics. Monolithic Microwave
Integrated Circuit (MMIC) designs were considered, but 
rejected based on cost and schedule concerns.
The FPA and RXB amplifiers in each frequency band are of identical 
design. In addition to requiring only one design per band, this also
provided for more flexibility in selecting and matching amplifiers 
in order to optimize radiometer performance.

Each amplifier has one drain voltage  supply, $V_d$, common
to all the transistors with nominal input voltages between +1.000~V 
and +1.500~V. Two gate bias supplies are provided on each
amplifier, one supplies gate bias voltage to the input stage, 
$V_\mathrm{gc}$, the other to all the remaining stages, $V_\mathrm{gf}$.
This biasing arrangement was chosen to allow for re-biasing 
the input stage of the cryogenic amplifiers for optimum noise 
performance, should the need arise. The externally
applied gate bias voltages range from 0.0~V to $-0.500$~V. 
Internal  resistive voltage dividers 
divide this voltage to that required by each device to 
compensate for variability in transistor characteristics.
Since cryogenic operation of HEMTs  may
require illumination, two light emitting diodes (LEDs) were designed into
each amplifier body. The amplifiers installed in the RXB have
both gate bias inputs  fed by a common voltage supply, since 
the ability to re-bias the input stage individually is not 
required, and the LEDs remain unpowered.

\subsubsection{Amplifier `Burn-in'}
During the construction of the first flight radiometers (Q1) 
it was observed that the drain current for a given gate bias and drain voltage
displayed a slow rise when operating at room temperature, with a
corresponding rise in amplifier gain. Similar effects are seen
with the K, Ka and V band amplifiers, all of which are
constructed using HEMT devices from a wafer which has a SiN  
passivation layer. The W-band amplifiers use HEMT devices
from an unpassivated wafer and do not exhibit this effect. The 
size of the effect is  largest for the small geometry devices
used in the higher frequency amplifiers, and seems to arise 
from a  thermally activated process. Once the amplifiers are cooled to
$\approx 100$~K the `burn-in' ceases. The largest effect is 
observed in  V-band, were the power gain, $g(t)^2$,
can be described approximately 
as $g^2(t) = g_0^2 (1 + \eta (1-e^{-t/\tau}))$ with 
$\eta \approx 1 $ and $\tau \approx 140$ hours. When
kept at room temperature and unpowered the amplifier's gain 
slowly returns to its initial state with a time constant of 
$\approx 500$ hours. 

During the radiometers' assembly the amplifiers were powered continuously, 
so the radiometers were constructed and characterized with the
amplifiers fully burned-in. Small changes in the insertion phase 
of the amplifiers associated with the gain changes,
could effect the in-flight  radiometer performance. The 
greatest concern relates to our knowledge of the frequency
dependent bandpasses of the radiometers. Such effects are 
expected to be small, since both amplifiers in a radiometer will
always have the 
same degree of burn-in. However, to ensure proper performance 
in flight, the radiometers remained powered as much as
practicable immediately before launch, since once launched the 
FPA temperature quickly falls preventing further burn-in of
the FPA amplifiers. Measurements are being performed on an 
engineering model of a V-band radiometer to estimate
the additional uncertainty in the knowledge of the radiometers' bandpasses,
 but it is not expected to be a significant effect.

\subsection{Phase-Matched Waveguides }
The phase-matched waveguides connect the output of the cold FPA HEMT amplifiers
to the inputs of the warm RXB amplifiers and were fabricated by Custom 
Microwave Inc.\footnote{Custom Microwave, Inc., 940 Boston Avenue, 
Longmont, CO 80501 USA}  Each waveguide incorporates 
$\approx 25$~cm straight section of commercially available 0.025~cm wall 
stainless steel waveguide
serving as a thermal break. Copper sections, overlapping 
$\approx 2.5$~cm at each end,  were electro-formed 
directly onto both ends of each stainless steel section, 
reducing the effective length of the thermal break
to $\approx$ 20~cm. Tellurium-copper waveguide flanges 
with over-sized exterior dimensions were soldered 
to the copper ends  of each waveguide assembly for transition 
to other waveguide components and structural support.
No plating or passivation treatment was applied to either 
the copper or stainless steel sections of the waveguides.
Insuring a \emph{broad-band} phase match required careful 
control of the dispersion relation of the waveguides.
Each phase-matched pair was designed
to have the same overall length and the same number and 
radii of `E' and `H' bends. The
electro-formed sections were fabricated on precision formed mandrels, that
were hand lapped to reduce deformations
in their cross section caused by bending. 
The dimensional tolerances of the commercially available stainless steel
waveguide sections were not well enough controlled to ensure 
a uniform dispersion relation from sample to sample, so
each stainless steel section was characterized, 
and matched sections delivered to Custom Microwave for 
incorporation into
the waveguide assemblies. To allow for manufacturing variations, the 
radiometer design used waveguide shims, 
up to one half  a guide wavelength, 
to be inserted at the warm end of one or both of the waveguide 
assemblies to adjust the electrical length of the assembly.  

\subsection{Phase Switches}
The phase switches  use a suspended stripline architecture and were manufactured by Pacific Millimeter Products.\footnote{
Pacific Millimeter Products, 64 Lookout Mountain Circle, Golden, CO, 80401 }
They can be set to two states differing in insertion phase by very nearly $180^\circ$, largely independent of frequency.
 Switching is accomplished by forward
biasing one of two back-to-back connected PIN diodes~\footnote{ Hewlett Packard  (HPND-4005) } through application
 of $\approx \pm 20$ \mA\ drive current to the
control input.  Table~\ref{table:phase_switch_specs} summarizes some of the important parameters of these devices. The housings were made of gold plated
aluminum, and the RF input and output connections were made with integrated waveguide to stripline transitions.
The PIN diode bias was applied using high speed constant current sources, switching polarity at 2.5 kHz. 
Precision matching of the average
insertion loss between the two states was obtained by adjusting
 the relative magnitudes of the positive and negative bias currents. 
As the drive current to the PIN diode is reduced, the  rate of change of the forward power
transmission coefficient with respect to drive current, $\Delta S_{21}/\Delta i$, increases.
For the purpose of estimating systematic errors from drive current variations, a conservative value of
 $\Delta S_{21}/ \Delta i = 0.03$/\mA\ was used, based
on measurements made with a PIN diode drive current of 5~\mA.  

\begin{table*}[t]
\caption{\small Flight Radiometer Phase Switch Specifications  }
\small{
\vbox{
\tabskip 1em plus 2em minus .5em
\halign to \hsize {#\hfil &\hfil#\hfil &\hfil#\hfil &\hfil#\hfil
                  &\hfil#\hfil &\hfil#\hfil \cr
\noalign{\smallskip\hrule\smallskip\hrule\smallskip}
{\sl MAP} Band Designation & K & Ka & Q & V & W\cr
Frequency range (\GHz) & 20-25  & 28-36 & 35-46 & 53-69  & 82-106 \cr
Insertion loss (dB) &$\approx 1.2$&$\approx 1.4$&$\approx 1.4$&$\approx 1.5$&$ \approx 2.5$ \cr 
Loss balance (dB)   &$\approx 0.1 $&$\approx 0.1$ &$\approx 0.1$ &$\approx 0.1 $&$\approx 0.3 $  \cr 
Phase difference (degrees)    &$   \pm 2   $&$\pm 2$      &$\pm 3$      &$\pm 3 $     &$\pm 3$        \cr
\noalign{\smallskip\hrule}
}}}
\label{table:phase_switch_specs} 
\end{table*}

\subsection{ Band Definition Filters}
The band  definition filters also used a suspended stripline technique, and were supplied by
Microwave Resources Inc.\footnote{Microwave Resources Inc., 14250 Central Avenue, Chino CA, 91710} They were manufactured with unplated aluminum bodies. Filter inputs and outputs are through waveguide ports, again implemented
 using integrated waveguide to stripline couplers. The band-edge frequencies and roll-off characteristics were specified
to limit the bandpass to frequencies where the input OMTs had acceptable return losses. 
Radiometric response where the OMT reflection 
coefficient becomes large could lead to large radiometric offsets.
 Table~\ref{table:filter_specs} summarizes the microwave performance
specifications of these filters.

\begin{table*}[t]
\caption{\small Flight Radiometer Band Definition Filter Specifications  }
\small{
\vbox{
\tabskip 1em plus 2em minus .5em
\halign to \hsize {#\hfil &\hfil#\hfil &\hfil#\hfil &\hfil#\hfil
                  &\hfil#\hfil &\hfil#\hfil \cr
\noalign{\smallskip\hrule\smallskip\hrule\smallskip}
{\sl MAP} Band Designation & K & Ka & Q& V & W \cr
$-1$~dB frequency (\GHz)    & 20.0, 25.0 & 28.5, 36.5  & 36, 44.5    & 54.5, 67.5 & 84, 104   \cr
$-15$~dB frequency (\GHz)    & 19.0, 26.0 & 27.0, 38.0  & 34, 46.5    & 51.5, 70.5 & 80, 108   \cr
Passband ripple (dB)&$\pm 0.25$   &$\pm 0.25$    &$\pm 1$      &$\pm 1 $    &$\pm 1 $   \cr
Insertion loss (dB) &$\approx 1$ &$\approx 1.2$&$\approx 2.4$&$\approx 2$ &$\approx 3$\cr
\noalign{\smallskip\hrule}
}}}
\label{table:filter_specs} 
\end{table*}

\subsection{Hybrid tees}
Both the FPA (cold) and RXB (warm) hybrid tees were manufactured by Millitech Corp.\footnote{Millitech,
 LLC, 29 Industrial Drive East, 
 Northampton, MA, 01060} The designs are based on Millitech's 
CMT series (90\% bandwidth version) hybrid tee. Changes made for {\sl MAP} include using 
aluminum bodies to reduce the mass, and, to eliminate the possibility of detached flakes, no gold plating.
A special epoxy, suitable for cryogenic operations, was used to secure the tuning post,
 and several screw holes were added and relocated
for mounting.
\subsection{Cold Hybrid Tee to OMT Waveguides}
These are the waveguide sections that connect the output of the OMTs to the inputs of the cryogenic hybrid tees. They
were fabricated from standard commercially available drawn copper waveguide and had waveguide flanges hard soldered to
their ends. The waveguide sections were bent and flanges attached by Microwave Engineering Corporation\footnote{ Microwave
Engineering Corporation, 1551 Osgood Street, North Andover, MA, 01845}. After fabrication they were annealed in a hydrogen
atmosphere to remove the work-hardening resulting from the forming process.
\subsection{Orthomode Transducers}
Orthomode transducers for all 5 frequency bands were designed and manufactured for {\sl MAP} by Gamma-f Corp.\footnote{ Now  
 Vertex RSI, 3111 Fujita Street, Torrance, CA 90505}
 All are
made of electro-formed copper and have very low insertions loss \citep{feedspaper}.
Table~\ref{table:omt_specs} summarizes some of the OMTs important performance characteristics. The design of the 
OMTs limit the usable microwave bandwidth of the radiometers.

\begin{table*}[t]
\caption{\small Flight Radiometer Orthomode Transducer Specifications  }
\small{
\vbox{
\tabskip 1em plus 2em minus .5em
\halign to \hsize {#\hfil &\hfil#\hfil &\hfil#\hfil &\hfil#\hfil
                  &\hfil#\hfil &\hfil#\hfil \cr
\noalign{\smallskip\hrule\smallskip\hrule\smallskip}
{\sl MAP} Band Designation & K & Ka & Q & V & W \cr
Frequency range (\GHz) & $20-25$ & $28-36$ & $35-46$  & $53-69$ & $82-106$ \cr
Return loss (dB)  &$ < -15$&$ < $-14$ $&$ < $-14$  $&$  < -15$&$ < -14$ \cr
Isolation   (dB)  &$ > 40 $&$ > $30  &$ > $30$ $ &$ > 27 $&$ > 25 $  \cr
Insertion loss~\tablenotemark{a}(dB)  &$ < 0.1 $&$ < 0.12$ &$ < 0.13$ &$ < 0.15 $ &$ < 0.22$      \cr 
\noalign{\smallskip\hrule}
}}
\tablenotetext{a}{The insertion loss values were measured at 300~K.}
}
\label{table:omt_specs}
\end{table*}

\subsection{Detectors}
The {\sl MAP} detectors were designed and supplied by Millitech Corp. They are a special design and include a tuning
element before the detector diode to lower their reflection coefficient, reducing the size of the standing wave
between the detector and the output of the RXB HEMT amplifier (Sec~\ref{sec:standing_waves}). In  order to design the line driver
circuit it was necessary to measure the detectors' video output characteristics. For output voltages up to 
$\approx 10$~mV the detectors have a very nearly square law response and can be modeled as a noiseless voltage
source and a series resistor in the $2000-3500~\Omega$ range. The resistor determines both the video output
 impedance of the 
detector and the Johnson voltage noise if it is assumed to be at $\approx300$~K physical temperature.
Table~\ref{table:det_specs} contains a number of the key performance parameters of the detectors. The responsivity
typically decreases with temperature by $\approx 0.01$~dB/K.

\begin{table*}[t]
\caption{\small Flight Radiometer Detector Specifications }
\small{
\vbox{
\tabskip 1em plus 2em minus .5em
\halign to \hsize {#\hfil &\hfil#\hfil &\hfil#\hfil &\hfil#\hfil
                  &\hfil#\hfil &\hfil#\hfil \cr
\noalign{\smallskip\hrule\smallskip\hrule\smallskip}
{\sl MAP} Band Designation & K & Ka & Q & V & W \cr
Frequency range (\GHz) & $20 - 25$ & $28 - 36$ & $35 - 46$ & $53 - 69$ & $82 - 106$ \cr
Return loss (dB max.)  &$  -3 $ &$ -5 $  &$  -5 $  &$   -2 $&$ -3    $\cr
Return loss (dB avg.)   &$  -6 $ &$-10 $  &$  -6  $ &$   -10 $&$ -7    $\cr
Responsivity (typical V/W) &$3000$ &$ 3500 $  &$2000$ &$ 2000 $&$ 2000 $\cr
Responsivity flatness (dB) &$\pm2$ &$ \pm 1.2 $&$\pm 1 $ &$ \pm 1.6 $
&$ \pm 2 $\cr
\noalign{\smallskip\hrule}
}}}
\label{table:det_specs}
\end{table*}

\subsection{Line Drivers}
The line drivers are x100  gain amplifiers  built on small circuit boards attached directly to
 the video output of the detector diodes. They boost the relatively small output voltages from the detector
 ($\approx 10$~mV) to a high level differential
signal, suitable for transmission over the instrument harness to the AEU. The line drivers 
were designed to limit the increase of the nominal operating detector
output noise by $< 0.5\%$,  and have enough dynamic range to ensure operation
 at room temperature (necessary  during integration and testing)
 without saturation. The input stage of the line driver is an Analog 
Devices\footnote{www.analogdevices.com} AD524 instrumentation amplifier.
 A protection circuit consisting of back-to-back Schottky diodes 
and a current limiting resistor was placed between the input of the AD524 and
 the output of the detector diode to protect the relatively
 sensitive detector diode  in the event that one of the bipolar power supplies to the line driver
 is absent. An additional Analog Devices OP37 operational amplifier
provides a differential output signal. The gain-bandwidth product of
 the instrumentation amplifier provides sufficient
 high frequency roll-off for out-of-band signals, so that no additional filtering is required. Fractional voltage gain
 variations due to
spin synchronous temperature and/or power supply voltage fluctuations are designed to be $< 0.5$~ppm 
and have no significant
contribution to the systematic error budget.

\section{ Support Electronics }
Careful design of the electronics supporting the radiometers is essential to
 achieve the desired radiometer performance. Of particular importance
are the thermal and supply voltage susceptibilities of the HEMT power supply
 circuits contained in the PDU and the science data
processing circuitry contained in the AEU. This section
 outlines the important design feature of {\sl MAP}'s instrument support electronics,
emphasizing techniques used to ensure the stable performance essential for the mission.
 Where appropriate, noise and stability requirements
derived from the systematic error analysis are presented . These are specified both as noise spectral power densities for random
noise and \rms\ values for spin synchronous terms. In general the limits for the random signals are derived from a requirement based on
 an increase in overall radiometer noise, whereas the spin synchronous terms are base on the systematic error budget.

\subsection{ The Power Distribution Unit (PDU) }
The PDU distributes power to all the instrument subsystems. It provides 31.5~V nominal spacecraft 
bus voltage to the AEU and DEU, and contains 5 switching mode DC-DC converters that power the FPA
HEMT amplifiers, RXB HEMT amplifiers, phase switch drivers, line drivers, and PDU commanding/housekeeping circuitry. (See Figure~\ref{fig:full_rad_dia}.)
The use of separate power converters for the different components
 was motivated by the need to eliminate
ground currents that are  potential sources of systematic errors.
 The 2.5~kHz phase switch drive signals,     
the switching frequencies of  the instrument DC-DC converters, and the sampling rate of the data collection
are all derived 
from a single crystal controlled
oscillator to eliminate the possibility of spurious time dependent signals caused by the beating of 
switching frequency leakage with the sampling frequency of the data collection system. 

 A great deal of attention
 was given to the elimination of ground loops
to ensure the safety of the HEMT amplifiers and to avoid ground loop induced radiometric artifacts. 

\subsubsection{FPA HEMT Bias Supplies}\label{sec:hemt_bias_supply}
	There are 40 separate bias supplies for the FPA HEMT amplifiers, all powered
 from a common switching power converter. The outputs
of the regulated switching DC-DC converter are fed into a set of linear
 voltage regulators, the outputs of which power the 
HEMT bias control circuity, effectively providing triple regulation of
 the bias voltages applied to the HEMT amplifiers. Each FPA HEMT
bias supply provides 4 outputs. One output supplies the drain bias voltage,
 and can be programmed to one of 8 equally spaced
voltage steps spanning $1.000$~V to $1.500$~V. The outputs are clamped
 at $-0.65$ and $+2.2$~V to protect the HEMT amplifiers from voltage transients
that could damage the amplifiers.
 The currents supplied by the drain bias supplies are monitored with $8.5~\mu$A resolution and are
sampled at 23.04 second intervals. Each FPA bias supply has two gate bias outputs
, one for the amplifier's
input stage, and one to supply the gate bias for all the following stages. Each of these
 voltages can be commanded from $0.000$ to 
$-0.500$~V in 16  steps. The gate bias supplies are clamped at $\pm 0.65 $~V for transient protection.
 The gate and drain bias supplies use separate sense and drive lines (including a remote ground sense line )
to compensate for voltage drops in the harness connecting to the
 HEMT amplifiers. The LED, used to illuminate the HEMT transistors, is driven
by a 5~\mA\ current source and has its own return line. The spin synchronous variation 
of the LED current is designed to be below 5~\nA.
 All the HEMT bias supplies float with respect to the ground of the PDU enclosure, and are
ground referenced through the harness to the cold FPA structure at the HEMT amplifier bodies. 
The harness is double shielded, with the
inner shield connected to the HEMT regulator ground at the PDU end only, and the outer
 shield connected to the PDU enclosure ground at both ends.
Table~\ref{table:hemtbias_noise} lists the noise specifications for these power supplies.

\begin{table*}[t]
\caption{\small HEMT Bias Supply Noise Specifications   }
\small{
\vbox{
\tabskip 1em plus 2em minus .5em
\halign to \hsize {#\hfil &#\hfil & #\hfil \cr
\noalign{\smallskip\hrule\smallskip\hrule\smallskip}
Requirement & Drain voltage supply & Gate voltage supplies \cr
\noalign{\smallskip\hrule\smallskip}
Broadband noise & $ <100~\nvolt~\rHz\;@2.5\pm 0.05~\kHz $ & $<100~\nvolt~\rHz\; @2.5\pm 0.05~\kHz$\cr
power spectral	& and first 10 harmonics  &  and first 10 harmonics  \cr
density         & $<23~\uvolt~\rHz,\;  1<f<$50$~\Hz  $  & $<20~\uvolt~\rHz,\; 1<f<50~\Hz$  \cr
& $<(23\cdot f~^{-0.45})~\uvolt~\rHz,  0.003<f<1~\Hz $ & $<(23\cdot f~^{-0.45})~\uvolt~\rHz, 0.003<f<1~\Hz$ \cr 
\noalign{\bigskip}
Spin synchronous& $<500~\nvolt~\rms @~f_\mathrm{spin}$ & 
$<400~\nvolt~\rms @~f_\mathrm{spin}$ \cr
variations      & $<500\cdot m^{1/2}~\nvolt \; @~m \cdot f_\mathrm{spin}$ & $<400\cdot m^{1/2}~\nvolt \; @~m \cdot f_\mathrm{spin}$\cr
\noalign{\bigskip}
Drift over mission & $< 10~\mvolt $& $<5~\mvolt$ \cr
\noalign{\smallskip\hrule}
}}
{\small m represents harmonic number associated with spin frequency, $f_\mathrm{spin}$ = .00773 Hz.} 
}
\label{table:hemtbias_noise} 
\end{table*}

The broad band noise specifications were derived from the requirement that the variance
 at the radiometers' outputs
resulting from broad band power supply noise be $< 1\%$ of the 
intrinsic radiometer noise, resulting in a 
$<0.5$ \% reduction in the radiometers' sensitivity. The requirements at bands
 centered at  2.5~kHz and harmonics arise since these are the 
frequency bands in which the radiometric information is contained.
 The spin synchronous specifications are derived from the systematic error allocation
 to the radiometers, which in turn are based on measured
 HEMT amplifier gain
variation coefficients derived from measurements of prototype amplifiers. Typical
values for the fractional power gain variation of the amplifiers, $(\Delta G/G)/\Delta V$, for small variations 
in the  drain  and gate voltages about their normal operating points are $1.1~V^{-1}$ and
 $1.2~V^{-1}$ respectively. 

\subsubsection{RXB HEMT Bias Supplies}  The 40 RXB HEMT bias supplies are  
similar to the FPA supplies and are powered from a separate secondary winding on the switching
power converter. Like the FPA
supplies, the RXB supplies have no ground reference to the PDU enclosure 
and are grounded solely by the connection to the RXB structure
through the HEMT amplifier bodies.   They each
have only one gate bias regulator that biases all the RXB amplifiers' gates, and there are
no current sources for the LEDs.

\subsubsection{ Phase Switch Driver Supplies} The phase switch driver supplies
 provide regulated $\pm 9.0$~V power to the 10 phase switch
driver circuit boards that are located in the RXB.

\subsection{ Phase Switch Drivers}
 Each phase switch driver circuit board
  powers the 4 phase switches on a DA, one jammed 
 and one toggled phase switch on each radiometer. All phase switches are driven
 by individually matched, high precision constant current
supplies, with current values in the 15-20 \mA\ range. 
The two toggled phase switches on each phase switch driver are operated 180 degrees out of phase
 (one supplies positive current
 while the other supplies negative current)  so as to minimize the modulated
 component of the current drawn from the power supplies.
 The jammed phase switches on each
assembly are also set to opposite polarity to balance the current draw from both polarity power
 supplies. As with the HEMT amplifier supplies,
the phase switch driver supplies in the PDU are isolated from the PDU enclosure
 ground, and are  ground referenced through the 
phase switch body's to the RXB structure. Common mode chokes are installed
 on the outputs of the phase switch drivers to reduce 
the size of any circulating currents through loops formed by the  coaxial
 cables attaching to the radiometer structure. 

Spin synchronous variations of the phase switch
drive current
 were specified to be less
that 1 \nA\ \rms. Based on the measured current dependence of the phase switch transmission coefficients,
this  keeps artifacts in the radiometric data from drive current variations below $0.3~\mu$K \rms.

\subsubsection { Line Driver Supplies } The PDU supplies doubly
 regulated $\pm$ 6.25~V to the line driver circuit boards that are attached
directly to the detectors on each radiometer. This supply  is also electrically isolated 
from the PDU structure, and receives its ground reference
through the line driver circuit boards. 

\subsubsection { Housekeeping and Interface Supplies } The housekeeping supply provides regulated $\pm$15V and +5V power to the control
 and current monitoring circuitry. 

 \subsection{The Analog Electronics Unit (AEU)} \label{sec:aeu}
\begin{figure*}
\epsscale{0.8}
\plotone{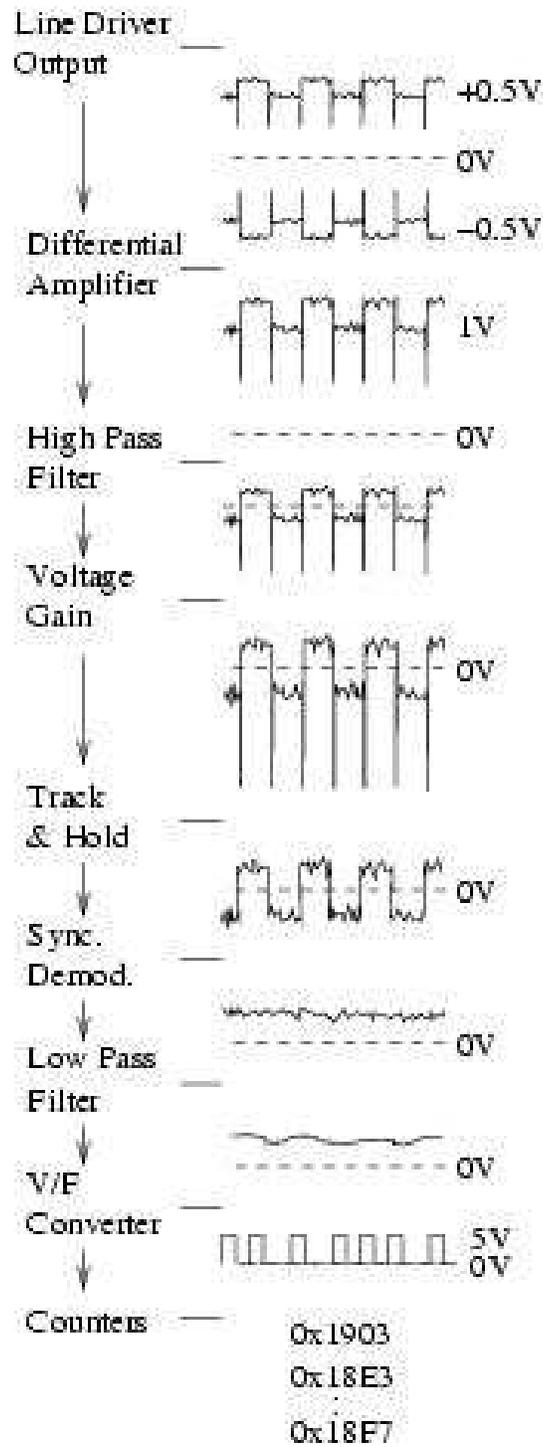}
\caption{Waveforms describing the stages of processing of a radiometer's signal in the AEU. The
 radiometric information is contained in the
amplitude and phase of the $2.5~\kHz$ square wave signal. The output of the V/F converter
is counted for $25.6~\ms$ intervals resulting in values centered at $\approx 6400$ counts. 
Details of the signal flow are described in
Section~\ref{sec:aeu}.}  
\label{fig:signal_flow}
\end{figure*}

The outputs of the 40 line drivers (2 for each radiometer) are amplified, filtered, demodulated,
digitized and integrated by the AEU,  and the digitized outputs are passed to the DEU. A block diagram
 outlining the functionality of the AEU is
 contained in Figure~\ref{fig:full_inst_dia} and representations of the radiometric signals at various
stages of processing are given in Figure~\ref{fig:signal_flow}.
\  Differential outputs from the
 line driver are received by a differential amplifier and converted to a single ended signal, 
referenced to the 
AEU analog ground. The DC component of this signal is sampled at  23.04 second intervals
 and digitized with  2.4~mV  resolution. This `RF bias'
signal provides  a monitor of the total RF power incident on the detectors, and is used to monitor the health of the radiometers. A simple RC
high pass filter with a 3~dB frequency of 8~Hz blocks the DC component, passing the radiometric data which
 appears as a 2.5~kHz square wave. The roll-off frequency of this
 filter was selected to
ensure that the response to  the 2.5~kHz signal varied by less than 1 ppm as a  result of
 the filter's phase shift variation arising from the temperature dependence of the capacitor. The 2.5~kHz
 output from the filter signal is amplified  and fed to the
 input of an Analog device AD630 synchronous
demodulator with its reference input clocked synchronously with the phase switch drive signal. The 
output of this demodulator is a
DC voltage, proportional to the temperature difference between the inputs of 
the radiometer (plus the radiometric offset) and an AC component
due to the radiometer's noise.

During each phase switch transition there is a several nanosecond period when
 both PIN diodes in the phase switch are off, briefly
reducing the RF  transmission through the phase switch. During this period the RF power
 incident on the detectors drops to roughly 1/2 its
normal value. Such a rapid change in the output voltage of the detector causes a brief 
period when the line driver outputs are slew-rate limited.
 Since the slew rates depend on the line drivers' power supply voltages and temperature, 
they could be a source of a systematic error
 had they been averaged into the radiometric data. Consequently, a  track and hold circuit 
blanks the output of the demodulator during the interval  from $1~\mu$s before the phase switch
 transition to $5~\mu$s after the phase switch transition, removing the transients from the radiometric data. 

The output of the track
 and hold circuit is filtered by  a two pole Bessel low pass filter with a 3~dB point of 100~Hz to 
bandwidth limit the signal. This signal is then sent to an Analog Device AD652 synchronous 
voltage-to-frequency (V/F) converter, clocked
at 1~MHz, resulting in a maximum output frequency of 500~kHz. Integration is performed by
 counting the output pulses of the converted signal for 25.6~ms intervals, providing a 0 - 12799 count
 digitization range. 

	Both the Bessel filter and the V/F converters introduce small correlations in the data.
The Bessel filter's response is described by its Laplace transform,
\begin{equation}
H(s) = \frac {A}{s^2 + Bs + A}
\end{equation}
where $ s = i \omega$,  $A = 6.803\times 10^5~s^{-2}$ and $ B = 1.360\times 10^3~s^{-1}$. This filter 
 negligibly affects the slowly
varying sky signal, but introduces  a $2.62~\%$ correlation of the white spectrum 
radiometer noise between adjacent 25.6~ms integrations.
 The synchronous
V/F converters incorporate first order sigma-delta modulators, for which the  
quantization noise in successive samples is anti-correlated.
 These anti-correlations can  manifest themselves as
idler-tones (limit cycle oscillations), resulting in significant spectral 
features in the quantization noise power spectrum. The
 idler tones quickly
diminish as the input noise level to the converters grows. The gain of the data collection 
system is set so that the noise
of the integrated output for each detector data channel is 2 counts
\rms\ or greater. Based 
on calculations and measurements, this level is
 sufficient  keep the quantization noise flat to $< 0.1$ counts.  All 40 channels of the data collection
system exhibited nearly identical responses. The noise covariance
 between adjacent 25.6~ms integrations can be 
approximated as
\begin{equation}
\overline{(y_i - \overline{y})  (y_{i+1} - \overline{y})} =  -0.09  + 0.0262\  \overline{(y_i-\overline{y})^2} 
\end{equation}
where the signal, $y_i$, is measured in counts.  The first term arises from the anti-correlated quantization introduced by the V/F converters,
 and the second term  from the response of the 
Bessel filter correlating the radiometric noise.

	The AEU contains its own power supplies consisting of a regulated DC-DC converter 
followed by a set of linear voltage regulators. It also contains the circuitry that reads the 
temperature of the 55 platinum resistance thermometers
which monitor the temperature of various instrument components. The thermometers are excited
 with an 195~Hz $200~\mu$A amplitude AC bias
 current and read out using
an automatically balancing  bridge circuit with a moving window.
 Each window allows for temperature measurement over an 8~K range with 0.5~mK resolution.
If the temperature read out falls outside the window boundary, the
 window's center temperature is moved by 4~K in the appropriate direction. The readout
 circuitry is also designed to ensure at least 1 count of random readout noise.
 This allows for averaging of repeated measurements to increase
the effective resolution of the thermometry. Each thermometer is read out at 23.04 second intervals.

\subsection{ The Digital Electronics Unit (DEU) }
	The DEU generates the the various control signals for the radiometers data 
processing and high resolution thermometers. It assembles
the radiometric data into packets, time tags them  and passes them to the data recorder for 
storage until they are down linked.
 The DEU also co-adds a number of successive
25.6~ms integrations for each radiometer to produce effective integration periods appropriate to
 the beam size of each radiometer. The beams on {\sl MAP}
move across the sky at approximately $2 \ddeg 6$/s, thus during each 25.6~ms integration period each
 beam moves by approximately $ 0 \ddeg 07$.  The number of  25.6~ms integrations co added for K, Ka, Q, V, and W
band radiometers are 5, 5, 4, 3, and 2 respectively.

\section{Assembly and Testing}
\subsection{Facilities}
The {\sl MAP} radiometers were assembled and initially tested in two clean rooms at Princeton University between 
 October  1997  and April 1999. The clean rooms were equipped with 4 test chambers of
 sizes ranging from 25 x 80 x 100~cm to
25 x 100 x 130~cm. The chambers each consist of a large rectangular aluminum plate 
with an  o-ring groove around the perimeter,
 and an aluminum clam-shell like cover. This design
provided good access to the radiometers when the cover was open, essential, since much of the final 
radiometer assembly took place while the 
radiometer was mounted in the chamber.
 The first stage of closed cycle refrigerator 
cooled a large aluminum `cold stage', on which the FPA components of the radiometer 
were attached,  to $\approx  90~K$. The second stage
of the refrigerator  cooled to approximately 15~K and provided both a cooling source 
for the  cryogenic loads, used to simulate
CMB signals, and  provided the cryo-pumping.

The warm (RXB) radiometer components
 were attached to a second large aluminum plate,
 the `warm stage', that
 was temperature controlled using Peltier heat pumps. The `warm stage' temperature could be adjusted from 
$-10$~C to + 50~C, and was used to vary the temperature of the warm components.

Test equipment available included vector network analyzers (VNAs), swept frequency sources and RF
 spectral analysis from 10~MHz to 170~\GHz, 
low frequency spectrum analyzers, and a set of large Helmholtz coils and power supplies, 
used to apply magnetic fields to the
radiometers for measuring magnetic susceptibilities.

  Each test
 chamber had a full set of power supplies,  used to power the
 radiometers, and a data collection system
used to log and analyze the radiometer's outputs. The design of
 these systems closely matched the design of the flight systems,
minimizing the risk of incompatibilities when the radiometers were
 connected to the flight electronics later in the program.
Lakeshore model 330 cryogenic temperature regulators were used to
 regulate the temperature of the cryogenic loads connected to the 
radiometers' inputs and the operating temperature of the 
FPA components on the `cold stage'.

\subsection{Procedure}
Components were visually inspected and characterized for
 key performance parameters, shown in Table~\ref{table:comp_char}. These data were used both to
 ensure that the components met performance specifications and as a basis for component selection.

\begin{table*}[t]
\caption{\small Component Characterization Summary  }
\small{
\vbox{
\tabskip 1em plus 2em minus .5em
\halign to \hsize {#\hfil & #\hfil \cr
\noalign{\smallskip\hrule\smallskip\hrule\smallskip}
Component & Parameters\cr
\noalign{\smallskip\hrule\smallskip}
Orthomode transducer        & $S_{21}$ ,main-arm $\rightarrow$ dual-mode, sidearm $\rightarrow$ dual-mode \cr
			    & $S_{11}$ ,main-arm, side-arm    \cr
Hybrid tee                  & $S_{21}$ ,colinear1 $\rightarrow$ E, colinear2 
$\rightarrow$ E,
 colinear1 $\rightarrow$ H, colinear2 $\rightarrow$ H,  E $\rightarrow$ H \cr
OMT-hybrid tee waveguides          & $S_{21}$, $S_{11}$    \cr
Phase-matched waveguides    & $S_{21}$, $S_{22}$, borescope  \cr
HEMT amplifiers             & $S_{21}$ (warm)   \cr
Phase switches              & $S_{21}$, $S_{11}$ (both states)  \cr
Band definition filters     & $S_{21}$, $S_{11}$   \cr
Detectors                   & $S_{11}$, responsivity, $Z_\mathrm{out}$  \cr
\noalign{\smallskip\hrule}
}}
{\small  Elements of the complex scattering matrix are $S_{xx}$.}
}
\label{table:comp_char}
\end{table*}

Using their measured characteristics, components were then selected to form the
two phase-matched legs of the radiometers. The input hybrid tees were selected for best E-port to H-port isolation, power balance
and loss balance, averaged across the radiometer's frequency band. The warm hybrid tees were selected based largely on power balance. Phase
switches to be toggled were selected for $S_{21}$ match between the two states, while those used in
the jammed leg were selected based upon the match of their insertion phase to a  corresponding toggled
phase switch. The complex gain of each phase-matched leg of the radiometer
 was then calculated based on the measured
component characteristics and various substitutions made to optimize the phase
 and amplitude matches. The phase-matched section
of the radiometer (all the components between the cold and warm hybrid tees inclusive)
 was then assembled in the test 
chamber with the selected components.

\begin{figure*}
\epsscale{1.3}
\plotone{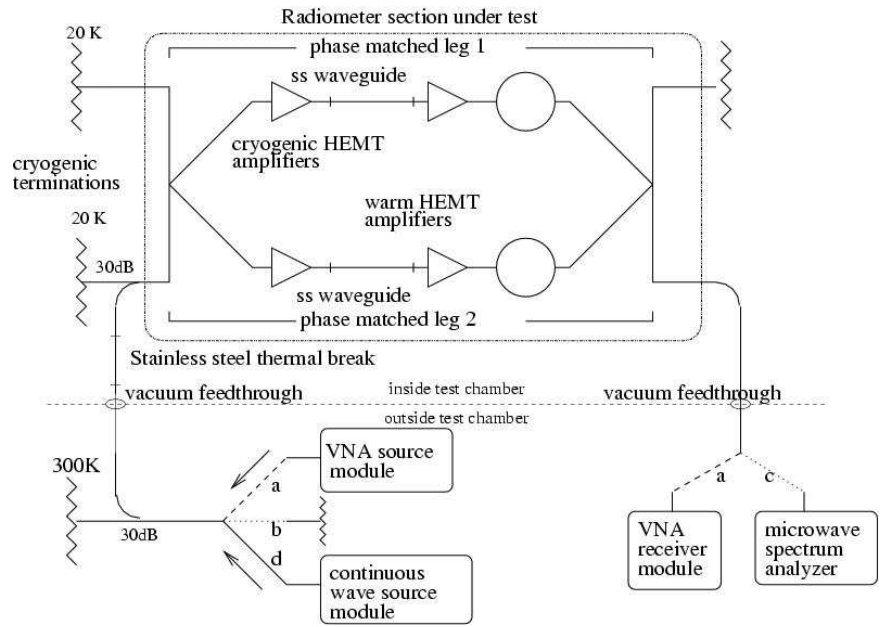}
\caption{ Test setup used to phase match the radiometers and to search for parasitic oscillations. For phase
matching the VNA source and detector modules were connected to the radiometer, shown by connections (a). Searches for 
in-band parasitic oscillation were performed with a load connected to the radiometer's input (b), and a spectrum
analyzer on the radiometer's output (c). Searches for out-of-band parasitic oscillations were performed with a continuous
wave source at the radiometers input (d), and a spectrum analyzer on the radiometer's output (c) }
\label{fig:phase_meas}
\end{figure*}

The VNA was connected to the partially assembled radiometer  as shown in Figure~\ref{fig:phase_meas}. 
 The signal from the source port of the VNA was fed into one of the input ports of the cold hybrid tee
after passing through two 30~dB couplers. These couplers were used to reduce the drive signal strength
 from the VNA so as not to saturate the warm HEMT amplifiers. To reduce
 the thermal noise level entering the radiometer during cryogenic tests, the 30~dB
coupler closest to the cold hybrid tee was located inside the test chamber, and operated
 at the same physical temperature as the other FPA
components. The main-arm of this coupler, and the unused input port of the cold hybrid tee were terminated
with waveguide terminations that could be operated at cryogenic temperature.
 One of the output ports of the warm hybrid tee
was attach to the receiving port of the VNA through  waveguide feedthroughs. The unused port of
 the warm hybrid tee was terminated with a room
temperature load.
By alternately powering the amplifiers in the two legs of the radiometers
it was possible to make accurate relative measurements of the complex gain of 
the two phase-matched legs. The phase and loss of the fixturing
components drops out when the ratio of the complex gains is taken, provided 
it remains constant over the several minute interval 
between measurements.

After verifying that the room temperature complex gain agreed with that
 modeled, the test chamber was pumped, leak tested and cooled, and
cold gain measurements performed. At this point small changes in amplifier
 bias were tried, to see if the overall phase and amplitude match
could be improved, while allowing for  a waveguide shim of up to 
 $\lambda_g/2$ in thickness to be inserted in either leg. Once an
optimum bias and shim thickness was determined, the radiometer was warmed 
 and the waveguide shim manufactured and installed. The room
temperature complex gain was then remeasured and the chamber cooled.
 The cold complex gain was then measured under
nominal operating conditions as well as with  variations in bias and 
 temperatures of the FPA and RXB components over the expected operating ranges. 

 Checks were made for parasitic oscillations by connecting a spectrum
analyzer to the port where the VNA receiver had been attached
 (See Figure~\ref{fig:phase_meas}.)  Measurements were then made with the source module of the VNA replaced
by a load to ensure no in-band parasitic oscillations were present.
 Additional measurements where made  with a continuous wave
 in-band  source connected
the the radiometer's inputs, to search for inter-modulation
 products resulting from an out-of-band oscillation that would not be directly
 observable due to the low frequency cutoff characteristics
 of the waveguide.

Upon completion of the preceding RF characterization, the radiometer was warmed,
 and a matched set of filters and detectors attached to the
two output ports of the warm hybrid tee, along with two temporary line drivers
 connected to the data collection system. Temperature regulated 
loads were attached to the two input ports of the hybrid tee, and the radiometer
 operated at room temperature as an initial check. The radiometer
was then cooled to nominal operating temperature and the temperatures of the two
 temperature regulated input loads set to the same value,
typically $\approx 20$~K. The levels of the phase switch bias currents were then
 adjusted to minimize the average amplitude of the 
square wave signal on the two detectors synchronous with the toggling of the phase
 switch, which effectively matched the band-average insertion loss
of the toggled phase switch in its two states. The insertion phase of
 the phase switches is essentially  independent of the drive
current over the ranges used, so there was no need to re-characterize
 the radiometers' phase match. 

The DC component
of the detector voltage was then measured and detector load resistors selected so that the in 
flight detector DC bias voltage

 would be about 10~mV. The load resistors are connected directly across the detector and have values
  ranging from none (open circuit) to $383~\Omega$,
with typical values of \mbox{$\approx 2{\rm~k}\Omega$}.
After  several preliminary measurements of the radiometer's offset, gain
and sensitivity, the radiometer was warmed, and the permanent resistors installed 
on the line driver circuit boards and phase switch driver
circuit boards. 

\section{Characterization}
Numerous test were performed to verify proper operation of the radiometers, supply
 a baseline for comparison with
later tests, and to measure some characteristics necessary for analysis of the flight data.
\subsection{ Basic Tests } A set of basic radiometric tests were performed
both with the radiometers at ambient temperature and at nominal operating temperature : 

 Gain Measurements -  The change in the demodulated radiometric
 output caused by changes in temperatures of test
loads applied to the radiometers' inputs. Each detector was calibrated versus each input.

Offsets - The temperature difference between the load temperatures required to bring the
 demodulated outputs from the
two detectors comprising a radiometer to the same level.

Noise - The radiometric noise spectral density  for each
 detector individually, and for
 the two detectors of each radiometer combined.

$f_\mathrm{knee}$ - The frequency at which the radiometer noise spectral density increases by a 
factor of  $\sqrt{2}$ above
its high frequency limiting value.

I-V Curves -  The I-V (current vs. voltage) characteristics of all the amplifiers were
 measured for each amplifier bias
input, while the other inputs for that amplifier remained at their nominal values.
 The value of the RF bias voltage
on each detector was also recorded during these tests. These data proved quite
 valuable for several applications.
First it provides a method for verifying the health of the  HEMT devices
 comprising the amplifiers. These devices are extremely
sensitive to over-voltages, such as produced by static discharges or
 turn-on transients of power supplies. Such damage almost always causes
a measurable change in the DC I-V characteristics of the device. These measurements
 therefore provide a good baseline with which to
 compare later measurements
to verify the health of the HEMT devices.

The detailed shape of the I-V characteristics of each amplifier is quite unique.
 This information was used during integration and
testing to verify that the entire end-to-end amplifier biasing system comprising software, bias
 supplies and harnessing was correctly
 configured for all 200 drain and gate voltage supplies. 

The detector RF bias values give a direct measurement of changes in the microwave gain of the radiometer
under different amplifier bias conditions. Such data has many uses, such as verifying that the
 amplifiers' gain versus
bias voltage coefficients are consistent with the values assumed in the systematic error budget. 

Finally, small discontinuities in the I-V characteristics (or their first derivative) often
 accompany the onset of parasitic
oscillations in microwave devices. Indeed, several such discontinuities associated with the onset
 of oscillations were
detected by these measurements, although not at the nominal in-flight operating biases.

\subsection{Basic Test Results}
Table~\ref{table:final_perf} presents a comparison of the radiometer sensitivity, offset and $f_\mathrm{knee}$
 frequency (defined in section \ref{1/fdef})
 as measured during assembly
and test, and during the second instrument level cold test at GSFC. The sensitivity
values presented are the best estimates of those expected in flight, and therefore have been scaled
to allow for the slight differences between the radiometer operating environment
 during testing and those expected in flight.
The Princeton measurements were made with reference loads attached directly to the input of 
the radiometers' OMTs, while the GSFC measurements
were performed with reference loads attached to the apertures of the feed horns. The differences
in $T_\mathrm{off}$ are believed to arise from the additional differential loss in the feed horns and
small reflections from the feed horn reference loads. 
The $f_\mathrm{knee}$ values in both cases are considered upper limits,
since the actual stability of the reference load temperatures is expected to be contribute to the observed value.

\begin{table*}[t]
\caption{\small Radiometer Performance Summary  }
\small{
\vbox{
\tabskip 1em plus 2em minus .5em
\halign to \hsize {\hfil#\hfil &\hfil#\hfil &\hfil#\hfil &\hfil#\hfil
                  &\hfil#\hfil &\hfil#\hfil  &\hfil#\hfil &\hfil#\hfil  \cr
\noalign{\smallskip\hrule\smallskip\hrule\smallskip}
Radiometer &
{\hskip0.5truein} $T_\mathrm{off} \ (\rm{K})$  &  & 
{\hskip0.5truein} $f_\mathrm{knee} \ (\Hz)$   & &
$\rm{Sensitivity} \ (\rm{mK~\sec}^{1/2})$ & &  
$\Delta T/{\rm pixel}\ (\ukelvin)$ \cr
 & PU & GSFC & PU & GSFC~\tablenotemark{a} & PU & GSFC 
& Flight prediction~\tablenotemark{b} \cr 
\noalign{\smallskip\hrule\smallskip}
K11  	& -0.03 & -0.033  &$<0.02$&$ <0.006$& 0.75 & 0.78&33.5\cr 
K12	& -0.04 & -0.249  &$<0.02$&$ <0.005$&0.82  & 0.94&\cr
Ka11	&  0.25 &  0.387  &$<0.02$&$ <0.002$&0.81  & 0.81&32.4\cr
Ka12 	&  0.05 &  0.072  &$<0.02$&$ <0.001$&0.81  & 0.83&\cr
Q11	&  0.34 &  0.092  &$<0.03$&$ <0.003$&1.06  & 0.95&31.5\cr
Q12	& -0.10 &  0.142  &$<0.03$&$ <0.003$&1.01 &  1.03&\cr
Q21	&  0.35 &  0.552  &$<0.01$&$ <0.002$&0.97  &  0.97&\cr 
Q22	&  1.60 & 1.036  &$<0.01$&$ <0.005$&1.13 & 1.13&\cr
V11	&  0.05 & -0.448  &$<0.03$&$ <0.003$&1.44 & 1.35&35.1\cr
V12	&  0.05 & -0.270  &$<0.03$&$ <0.004$&1.12 & 1.16&\cr
V21	& -0.80 & -0.265  &$<0.03$&$ <0.002$&1.13 & 1.09&\cr
V22	&  0.02 &  0.352  &$<0.02$&$ <0.003$&1.16 & 1.23&\cr
W11	&  0.28 & -0.451  &$<0.03$&$ <0.016$&1.59 & 1.29&30.7\cr
W12	& -1.63&-2.064  &$<0.02$&$ <0.015$&1.78 & 1.53&\cr
W21	&  0.13 & -0.091  &$<0.01$&$ <0.002$&1.79 & 1.50&\cr
W22	& -0.35 & 0.008  &$<0.01$&$ <0.001$&1.81 & 1.57&\cr
W31	&  0.35 &-1.151  &$<0.01$&$ <0.009$&1.74 & 1.60&\cr
W32	&  0.18 &-1.117  &$<0.01$&$ <0.021$&1.94 & 1.84&\cr
W41	& -0.58 & 1.300   &$<0.015$&$ <0.008$&1.68& 1.73&\cr
W42     & -1.12& 1.441  &$<0.02$&$ <0.005$&1.70 & 1.56&\cr
\noalign{\smallskip\hrule}
}}
{\small Summary of the $T_\mathrm{off}$, $f_\mathrm{knee}$,
and sensitivities of the 20 radiometers comprising {\sl MAP} as measured during radiometer
 construction (Princeton) and during integration and testing (GSFC). The sensitivity values given are for the 
combined output of the two detectors
on each radiometer. The sensitivity values have been scaled to approximate those 
expected in-flight for a projected FPA temperature of 95~K. Differences 
in $T_\mathrm{off}$ and sensitivity
are attributed to the different input loads and the additional loss from the feed horns in the GSFC test.
Actual in-flight performance will be slightly different.}
\tablenotetext{a}{The entries in this column are the measured values of $f_\mathrm{knee}$  observed  during
the final instrument level cold test. They include contributions from both the radiometer instabilities
and temperature fluctuations of the feed horn aperture loads and are therefore be interpreted as upper limits
in terms of radiometer performance.}
\tablenotetext{b}{The values given are the predicted band combined $1\sigma$  Rayleigh-Jeans temperature uncertainties
per pixel calculated from the GSFC test results. These values assume $3.2\times 10^{-5}$ sr pixels, uniform sky coverage
 and a nominal 2 year observing period. } 
}
\label{table:final_perf}
\end{table*}

\subsection{Additional Tests}
A number of additional tests were performed only with the radiometers near their
 nominal in-flight operating temperature. 

\subsubsection{ Warm and Cold End Temperature Susceptibility Coefficients}  One possible source of systematic error
 results from  changes in a radiometers' performance caused by small, periodic temperature variations. Much
 effort was expended to limit both the 
size of such temperature variations and to minimize the associated change in the radiometers' characteristics.
 Susceptibility coefficients
were measured by monitoring the radiometers' outputs as the temperatures of the warm and cold sections
 were varied over a 10-20~K range,
in order to produce measurable effects.
 Susceptibility
coefficients are required to relate the change in radiometer performance to the change in
 the physical temperature of the radiometer components.
 Measurements were made of the variations of offset and gain  versus the
 temperature of both the FPA and RXB sections. 

Offset versus 
temperature  measurements were made by setting the temperature of the two
input loads to null the radiometers' outputs. The temperature of each radiometer 
section (FPA or RXB) was then varied, and the 
resulting changes in the radiometers' outputs recorded. Since
 the radiometers' outputs were initially nulled gain changes have little effect on
 the radiometers' outputs, so the offset dependence is easily determined.
Gain variations were measured by setting a  $\approx 10 - 20$~K  temperature
 difference between the input loads, varying the temperature
of the radiometers' sections and recording the radiometers' output. 
These data have effects both from the temperature dependence of the offset
and the varying gain, however, since the offset dependence was measured separately 
it is easily removed, allowing for determinations of the 
gain vs. temperature coefficient.

 These coefficients, combined with the predicted  temperature variations, were used to verify that the
expected in-flight radiometer stability met specification.  In flight, the actual temperature variations of 
55 key components are
 measured with 0.5~mK resolution, allowing for a direct measure of the thermal stability of the radiometers, 
verifying both the thermal design
and, with use of the aforementioned susceptibility coefficients, the size
 of expected temperature induced radiometric artifacts.
 No corrections to the radiometers' outputs are expected to be necessary due to
temperature variation induced effects in the radiometers. 
\subsection{ Observatory Level Tests}
After the instrument was mated to the observatory tests were performed to search for
interactions between the instrument and spacecraft systems. The final observatory level
thermal vacuum test was conducted with the instrument operating near temperatures expected in flight
and with the cryogenic temperature regulated loads attached to the feed horn apertures. The outputs
 of the radiometers were recorded as the spacecraft systems were exercised.
Tests included searches for radiometric artifacts  induced by the reaction wheels,
 transponder, bus voltage fluctuations and
radio frequency noise coupled to the instrument through the power bus. No unexpected effects
 were observed during these tests
and  no corrections to the flight radiometric data arising 
from interactions with the spacecraft systems are anticipated.

\subsubsection{ Bandpass Measurement} 
\begin{figure*}
\epsscale{1.4}
\plotone{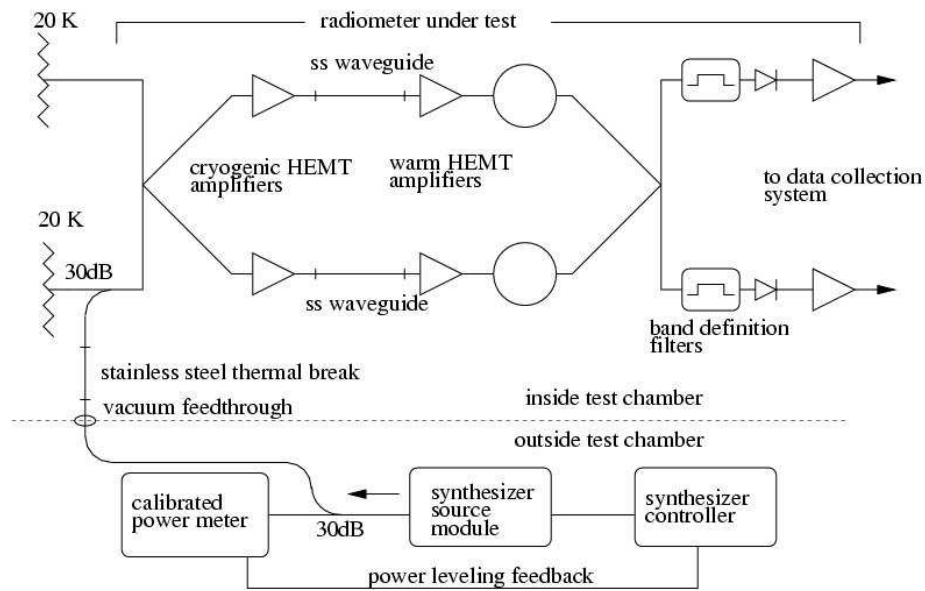}
\caption{Test setup used to measure the radiometer's bandpasses. The calibrated power meter and synthesizer
are configured to provide a constant power at the output of the room temperature 30~dB coupler. The parameters
of this coupler, and all the fixturing waveguide used to route the signal to the cold hybrid tee
 at the radiometer's
input were characterized with the VNA, and corresponding corrections applied to the measured bandpass data. 
 \label{fig:bandpass_meas} }

\end{figure*}

Measurements of the radiometers' bandpasses are necessary for the 
analysis of in-flight data. Unlike the radiometric
gains and offset, which are calibrated in flight, bandpass measurements must
 be made before launch. Measurements were made by injecting
a small signal into the one port of the cold hybrid tee and recording the radiometric response.
 The level of the signal was
set to yield  a readily measurable radiometric response, but not so large as to cause
 non-linear effects. Figure~\ref{fig:bandpass_meas} shows the test
configuration used for these measurements. The VNA was used to make precise measurements
 of the frequency dependent transmission characteristics
of all the components between the leveled source and radiometer's input at the cold hybrid
 tee. The swept signal source was 
calibrated to produce a frequency independent power level at the output of the through
 port of the room temperature 30~dB coupler. The bandpass was  measured by
stepping the synthesized source through 201 frequencies, $\nu_i$, and recording the
 output of the radiometer at each frequency with the RF output of the 
synthesizer enabled and disabled. Differences between the data taken with the synthesizer
 on and off then yielded the radiometer's response to the applied
signal. Corrections were applied to these data to account for the frequency
 dependence of couplers and waveguide. Measurements were made at different power levels to ensure
 that non-linear effects arising from the large signal levels
were not present. 

The procedure resulted in a bandpass measurement relating the response
from each detector
to the input microwave \emph{power}, referenced to the input port of the
cold hybrid tee, so it does not include possible frequency dependences of the feed horn, OMT
 and waveguides connecting the OMT to the cold hybrid tees. 
The transmission coefficients of the OMT and waveguide sections were measured with the VNA
 and small ($ <0.3$~dB) corrections applied
to the measured responses, resulting in corrected responses,
 $r_l(\nu_i)$ and $r_r(\nu_i)$, for each detector. 

\begin{figure*}
\epsscale{1.5}
\plotone{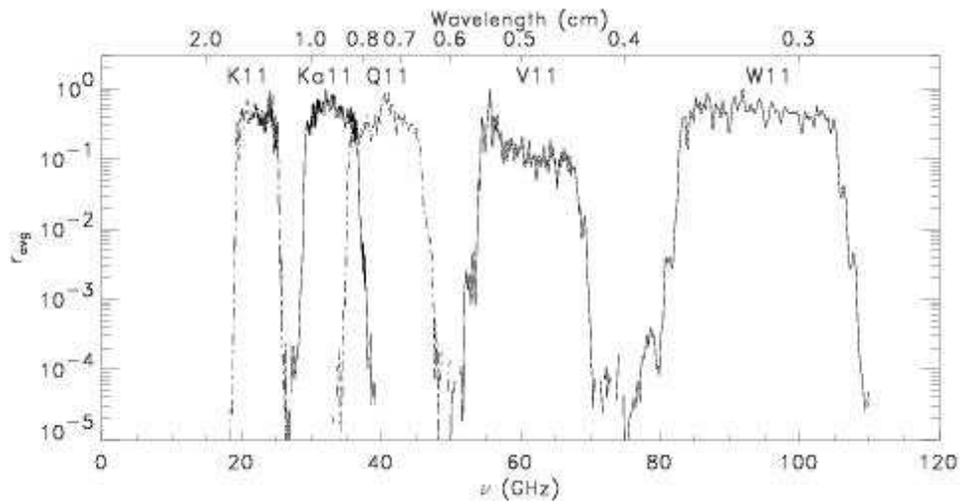}
\caption{ Combined responses, $r_{\mathrm{avg}}(\nu)$, 
of  representative radiometers in each of
{\sl MAP's} five frequency bands. Note that the frequency 
ranges of the Ka and Q band radiometers
overlap slightly.
 \label{fig:bandpasses} }
\end{figure*}

Absolute calibration of the radiometers is performed in flight using observations of the CMB dipole. The
demodulated digitized output of each detector, $y_j$, can be expressed as
\begin{equation}
y_j = y_{\mathrm{off},\, j} + y_{0,\, j} \sum_i r_j(\nu_i) w_{A-B}(\nu_i) 
\end{equation} 
where $j = l,\, r$ designates the detector of the radiometer,
$y_{\mathrm{off},\, j}$ is the radiometric offset, 
 $y_{0,\, j}$, is the overall gain of the radiometer and data collection system, and
$w_{A-B}(\nu_i)$, is the power per unit bandwidth at frequency $\nu_i$ of the differential signal
coupled to the radiometer's input. Calibration is performed by comparing the \emph{change} 
in each detector's output signal,
$\Delta y_j$,
to the corresponding change in the input differential power signal arising from the CMB dipole :
\begin{eqnarray}
\Delta y_j&=&y_{0,\,j} \left(\sum_i r_j(\nu_i) w^\prime(\nu_i)\right) \Delta(T_A - T_B) \\
w(\nu)&=&\frac{h \nu}{e^x- 1}\\
x&=&\frac{h \nu}{k_B T}\\
w^\prime(\nu)&\equiv&\left. \frac{1}{k_B}\frac{dw(\nu)}{dT}  \right|_{T = T_\mathrm{CMB}} 
=  \frac{x^2 e^{x}}{(e^{x} - 1 )^2}. 
\end{eqnarray}
Here $w(\nu)$ is the power per unit bandwidth delivered to a single mode by a blackbody
at temperature $T_{CMB} = 2.725$~K, $h$ is Planck's constant, and $\Delta(T_A - T_B)$ is the change in the 
differential  thermodynamic temperature of the observed CMB dipole calculated
from the orientation of the satellite. The calibration constants, $\widehat{r}_j$,  are defined so that
digitized data from each detector may be converted into measured thermodynamic temperature differences,
 $\Delta \widehat{T}_{\mathrm{\mathrm{th}},\, j}$, by
\begin{equation}
\Delta \widehat{T}_{\mathrm{\mathrm{th}}, \, j} = \widehat{r}_j \Delta y_j.
\end{equation}
The calibration constants are 
measured using the relation
\begin{eqnarray}
\widehat{r_j}^{-1} &=& \left<\frac{\Delta y_j}{\Delta(T_A - T_B)}\right>  \\
&=& y_{0,\,j} \sum_i r_l(\nu_i) k_B w^\prime(\nu_i)  
\end{eqnarray}
where the brackets indicate an average over many observations. 
 The calibrated data of the two detectors are combined with a simple average to yield the
calibrated radiometer output, 
\begin{equation}
\Delta \widehat{T}_{\mathrm{th},\, \mathrm{avg}} =\frac{1}{2}( \widehat{r}_l \Delta y_l + \widehat{r}_r \Delta y_r). 
\end{equation}
The frequency response of this combined signal is an average of the two detector
frequency responses
\begin{equation}
r_{\mathrm{avg}}(\nu_i) =
 \frac{1}{2}\left(\frac{r_l(\nu_i)}{\sum r_l(\nu_i) w^\prime(\nu_i)} +
\frac{r_r(\nu_i)}{\sum r_r(\nu_i) w^\prime(\nu_i)}\right). 
\end{equation}
Combined responses of representative radiometers in each of MAP's frequency bands are presented in
Figure~\ref{fig:bandpasses}.

The response of a radiometer to a signal with an arbitrary spectrum is obtained by integrating
the product of $r_{\mathrm{avg}}(\nu)$ and the power spectrum over the passband of the radiometer.
The factors used to convert thermodynamic temperature differences to Rayleigh-Jeans temperature
differences for the calibrated radiometer output 
were calculated using the relation
\begin{equation}\label{eqn:Tth_to_TRJ}
\frac{d(\Delta \widehat{T}_{\mathrm{th}, \,\mathrm{avg}})}{dT_{\mathrm{RJ}}} =
\frac{\sum r_{\mathrm{avg}}(\nu_i)}{\sum r_{\mathrm{avg}}(\nu_i)w^\prime(\nu_i)}
\end{equation}
and are listed in Table~\ref{table:bandpass_coeff}. 
The numerator on the right hand side of equation~\ref{eqn:Tth_to_TRJ} is uniformly weighted with
frequency since the power per unit bandwidth in  Rayleigh-Jeans units is independent of frequency.  
The thermal center frequency for each radiometer, $\nu_{\mathrm{th}}$, is the frequency
at which the Rayleigh-Jeans
to thermodynamic conversion equals that calculated for the radiometer, 
\begin{equation}
\frac {d(\Delta \widehat{T}_{\mathrm{th},\,
\mathrm{avg}})}{dT_{\mathrm{RJ}}} =
\frac{1}{w^\prime(\nu_{\mathrm{th}})}
\end{equation}
and are also given in Table~\ref{table:bandpass_coeff}.
The factors
 need to convert an observed Rayleigh-Jeans temperature difference to a Rayleigh-Jeans temperature
 difference at 
another frequency, $\nu_0$, for objects with antenna temperature scaling as $T_A\propto \nu^{\beta}$ were
 calculated from the relation
\begin{equation}
\frac{dT_{\mathrm{RJ}}}{dT_{\mathrm{RJ}}(\nu_0)}(\beta) =\frac{\sum  r_{\mathrm{avg}}(\nu_i)(\frac{\nu_i}{\nu_0})^\beta}
{\sum r_{\mathrm{avg}}(\nu_i)}. 
\end{equation}
In these units a source with a constant Rayleigh-Jeans temperature has $\beta = 0.$
The resulting coefficients were fit to a function of the form
\begin{equation}
\frac{dT_{\mathrm{RJ}}}{dT_{\mathrm{RJ}}(\nu_0)}(\beta) = \left(\frac{\nu_{\mathrm{avg}}}{\nu_0}\right)^\beta( 1 + c_1\beta + c_2\beta^2)
\end{equation}
for $ -3 < \beta < 1 $ 
and the resultant parameters, $\nu_{\mathrm{avg}}$, $c_1$ and $c_2$ are  given in Table~\ref{table:bandpass_coeff}. 

 These quantities apply to 
diffuse sources that completely fill the beams of the instrument. For resolved sources
 additional terms resulting from the 
frequency dependent beam sizes must be included \citep{opticspaper}. 
Tabulated bandpass measurements will be made available as part of the  {\sl MAP} data release.

\begin{table*}[t]
\caption{\small {\sl MAP} Radiometer Bandpass Parameters  }
\small{
\vbox{
\tabskip 1em plus 2em minus .5em
\halign to \hsize {\hfil#\hfil &\hfil#\hfil &\hfil#\hfil &\hfil#\hfil
                  &\hfil#\hfil   &\hfil#\hfil &\hfil#\hfil  \cr
\noalign{\smallskip\hrule\smallskip\hrule\smallskip}
Radiometer & $\nu_{\mathrm{avg}} \ (\GHz)$ & $c_1$ & $c_2$ &
$\Delta \nu_{eff}\  (\GHz)$ & $\nu_{\mathrm{th}} (\GHz)$ &
$dT_{\mathrm{th}}/dT_{\mathrm{RJ}}$\cr
\noalign{\smallskip\hrule\smallskip}
K11&   22.29 &   -2.43E-03 &  2.78E-03 & 5.26  & 22.36 & 1.013 \cr
K12&   23.12 &   -2.35E-03&   2.83E-03 & 4.09  & 23.18 & 1.014\cr
Ka11&  32.78 &   -1.58E-03&   1.75E-03 & 6.75  & 32.84 & 1.028\cr
Ka12&  33.12 &   -1.85E-03&   2.06E-03 & 7.04  & 33.19 & 1.029\cr
Q11&   40.71 &   -1.85E-03&   2.10E-03 & 8.62  & 40.79 & 1.044\cr
Q12&   40.81 &   -1.48E-03&   1.70E-03 & 7.66  & 40.88 & 1.044\cr
Q21&   40.16 &   -1.58E-03&   1.78E-03 & 7.55  & 40.23 & 1.043\cr
Q22&   40.92 &   -1.53E-03&   1.78E-03 & 7.68  & 40.99 & 1.044\cr 
V11&   59.20 &   -2.13E-03&   2.20E-03 & 8.37  & 59.32 & 1.095\cr
V12&   61.10 &   -2.00E-03&   2.23E-03 & 12.53 & 61.22 & 1.101\cr
V21&   61.61 &   -1.95E-03&   2.22E-03 & 12.81 & 61.72 & 1.103\cr
V22&   60.63 &   -2.12E-03&   2.35E-03 & 11.66 & 60.75 & 1.099\cr
W11&   93.60 &   -2.02E-03&   2.20E-03 & 20.81 & 93.71 & 1.249\cr
W12&   93.18 &   -1.69E-03&   1.82E-03 & 18.30 & 93.27 & 1.247\cr
W21&   93.49 &   -1.47E-03&   1.57E-03 & 16.54 & 93.58 & 1.248\cr
W22&   94.27 &   -1.37E-03&   1.48E-03 & 16.54 & 94.35 & 1.253\cr
W31&   92.36 &   -1.35E-03&   1.44E-03 & 16.70 & 92.44 & 1.242\cr
W32&   93.32 &   -1.30E-03&   1.37E-03 & 16.15 & 93.39 & 1.247\cr
W41&   94.24 &   -1.67E-03&   1.83E-03 & 18.76 & 94.33 & 1.253\cr
W42&   93.09 &   -1.58E-03&   1.68E-03 & 17.73 & 93.18 & 1.246\cr
\noalign{\smallskip\hrule}
}}}
\label{table:bandpass_coeff}
\end{table*}

\section{Summary}The design, assembly, and test procedures for the 20 differential microwave radiometers
comprising {\sl MAP} have been described. Through the use of rapid phase switching and instantaneous differencing 
radiometer instabilities arising from the HEMT amplifiers' gain fluctuations have been dramatically reduced. 
 Ground characterization tests indicate that the MAP radiometers and support electronics
will meet MAP's  sensitivity and systematic error goals when operating in flight.  

\section{Acknowledgments}
The {\sl MAP} radiometers are the result of countless hours of work by many dedicated individuals.
N. Bailey, T. Boyd, R. Harris, W. Lakatosh,  D. Thacker, 
J.Webber , and B. Wireman at NRAO produced the remarkable
HEMT amplifiers which are the heart of this project. 
The National Radio Astronomy Observatory is a facility of the National
Science Foundation operated under cooperative agreement 
by Associated Universities, Inc. 
A. Hislop of Pacific Millimeter Projects and H. Arain
of Microwave Resources Inc. worked closely with {\sl MAP} to produce the phase switches and filters.
 D. Bergman, D. Brigham, J. Caldwell, and C. Kellenbenz of  NASA/GSFC
were responsible for the radiometer support electronics.
 M. Jones
and M. Delmont of NASA/GSFC provided essential quality assurance oversight to Princeton, NRAO and numerous 
component vendors. C. Jackson was invaluable as instrument system engineer in coordinating the
efforts of the different groups. At Princeton
technical support was provided by  R. Bitzer, W. Groom,  R. Sorenson, and C. Sule.  G. Atkinson,
W. Dix, J. Mellodge and L. Varga in the Princeton machine shop worked on both the radiometer 
test facilities and the flight hardware.
 Administrative and purchasing
services at Princeton were provided by S. Dawson, H. Murray, A. Qualls, and K. Warren. This research
 was supported by the
 {\sl MAP} project under the NASA Office of Space Science and Princeton University. 

\appendix
\section{ Detector Noise Covariance } \label{app:noise_cov}
\begin{figure*}[t]
\epsscale{.15}
\plotone{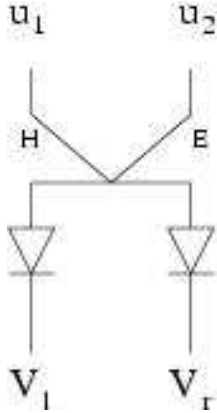}
\caption{Noise sources and detectors attached to a hybrid tee. E and H are labels for the arms of the hybrid tee.}
\label{fig:noise_cov}
\end{figure*}

Consider the  situation where two uncorrelated voltage noise sources, $ u_1$ and $u_2$,  are connected to the
inputs of a hybrid tee, and two square law detectors are connected to the output as shown in
 Figure~\ref{fig:noise_cov}. The voltage noise at the input to the two detectors becomes
\begin{equation}
\frac{u_1 + u_2}{\sqrt{2}}\ \textrm{and}\ \frac{u_1 - u_2}{\sqrt{2}}.
\end{equation}
The output voltages of the two square law detectors are proportional to
the squares of these quantities, with proportionality
constants $s_l$ and $s_r$ for the two detectors.
The covariance between the two detector outputs is
\begin{eqnarray}
\overline{(V_l - \overline{V_l})(V_r - \overline{V_r})}& = & \left(\overline{(V_lV_r)} - \overline{V_l} \ \overline{V_r}\right)\\
&=&\frac{ s_l s_r}{4}\left(\overline{(u_1+u_2)^2(u_1-u_2)^2} - \overline{(u_1^2 + u_2^2)}^2\right)\\
&=&\frac{ s_l s_r}{4}\left((\overline{u_1^4} + \overline{u_2^4} - 2 \overline{u_1^2u_2^2} - (\overline{u_1^2} + \overline{u_2^2})^2\right).
\end{eqnarray}

For a normally distributed variable, such as $u_i^2$,   we have $ \overline{u_i^4} = 3 \overline{u_i^2}^2 $, so
\begin{equation}.
\overline{(V_l - \overline{V_l})(V_r - \overline{V_r})} 
= \frac{ s_l s_r}{4} 2(\overline{u_1^2}^2 +\overline{u_2^2}^2 -2\overline{u_1^2u_2^2})
= \frac{ s_l s_r}{2}\left(\overline{u_1^2} - \overline{u_2^2}\right)^2 
\end{equation}

\pagebreak


\begin{thebibliography}{}
\bibitem[Barnes(2002)]{feedspaper}Barnes et al. 2002, \apjs, 143, 567
\bibitem[Bennett et al.(2003)]{missionpaper}Bennett et al. 2003,  
\apj, 583, in press
\bibitem[Dicke(1946)]{dicke}Dicke, R.H. 1946, Rev. Sci. Instrum., 17, 268
\bibitem[Jarosik(1996)]{corrnoisepaper}Jarosik, N.C., IEEE Transactions on Microwave Theory and Techniques 1996, 44, 193
\bibitem[Kogut et al.(1996)]{cobepaper}Kogut et al. 1996, \apj, 470, 653
\bibitem[Kraus (1986)]{Kraus}Kraus, J.D. 1986, Radio Astronomy 2nd Ed., Powell, Ohio, Cygnus-Quasar Books
\bibitem[Page(2003)]{opticspaper}Page et al. 2003 \apj, 585, in press
\bibitem[Pospieszalski(2000)]{amplifierpaper}Pospieszalski, M. P., Wollack, E.J., Bailey, N., Thacker, D., Webber, J. 2000,
 IEEE MTT-S International Microwave Symposium Digest, Boston, MA, 1, 25
\bibitem[Predmore(1985)]{predmore}Predmore, C.R. et al. 1985, IEEE Transactions on Microwave Theory and Techniques, 33, 356
\bibitem[Wedge(1992)]{wedge} Wedge, S.W. \& Rutledge, D.B. 1992,  IEEE Transactions on Microwave Theory and Techniques, 40
\bibitem[Wollack(1998)] {1/fpaper}Wollack, E.W. \&  Pospieszalski, M.W., Proc. 1998 IEEE MTT-S Int. Microwave Symp. Digest,
Baltimore, MD, p 669  

\end{thebibliography}
\end{document}